\documentclass[manuscript]{acmart}
\usepackage{longtable}
\usepackage{csquotes}
\usepackage[nolist,nohyperlinks]{acronym}
\usepackage{graphicx}
\usepackage{natbib} 
\usepackage{makecell}
\usepackage{adjustbox}
\usepackage{tabularx}
\usepackage{booktabs} 
\usepackage{array}    
\usepackage{caption}  
\usepackage{multicol}
\usepackage{multirow}
\usepackage{spverbatim}
\graphicspath{ {./figures/} }

\begin{acronym}
    \acro{AI}{Artificial Intelligence}
    \acro{GenAI}{Generative Artificial Intelligence}
    \acro{QRS}{Qualitative Research Synthesis}
    \acro{HCI}{Human-Computer Interaction}
    \acro{PCG}{Procedural Content Generation}
    \acro{LLM}{Large Language Model}
    \acro{GPT}{General Pre-Trained Transformer}
    \acro{TTIG}{Text-to-Image Generation}
    \acro{eMERGe}{enhancing Meta-ethnography Reporting Guidelines}
    \acro{CASP}{Critical Appraisal Skills Programme}
    \acro{PRISMA-S}{Preferred Reporting Items for Systematic Reviews and Meta-Analyses literature search extension}
\end{acronym}


\usepackage{xcolor}
\definecolor{Lgroup}{HTML}{5A56D2} %violet-blue
\definecolor{Pgroup}{HTML}{00875B} % teal-green
\newcommand{\Lgroup}[1]{{\color{Lgroup} \emph{#1}}}
\newcommand{\Pgroup}[1]{{\color{Pgroup} \emph{#1}}}

\title{Generative AI in Game Development: A Qualitative Research Synthesis}

\author{Alexandru Ternar}
\orcid{0009-0006-6917-0675}
\affiliation{%
  \institution{Aalto University}
  \department{Department of Computer Science}
  \city{Espoo}
  \country{Finland}
}

\author{Alena Denisova}
\email{alena.denisova@york.ac.uk}
\orcid{0000-0002-1497-5808}
\affiliation{%
  \institution{University of York}
  \department{Department of Computer Science}
  \city{York}
  \country{UK}
}

\author{Jo\~ao M. Cunha}
\email{jmacunha@dei.uc.pt}
\orcid{0000-0001-7216-8396}
\affiliation{%
  \institution{University of Coimbra}
  \department{CISUC/LASI, DEI}
  \city{Coimbra}
  \country{Portugal}
}

\author{Annakaisa Kultima}
\email{annakaisa.kultima@aalto.fi}
\orcid{0000-0001-5856-5643}
\affiliation{%
  \institution{Aalto University}
  \department{Department of Art \& Media}
  \city{Espoo}
  \country{Finland}
}

\author{Christian Guckelsberger}
\email{christian.guckelsberger@aalto.fi}
\orcid{0000-0003-1977-1887} 
\affiliation{
  \institution{Aalto University}
  \department{Department of Computer Science}
  \city{Espoo}
  \country{Finland}
}
\affiliation{
  \institution{Queen Mary University of London}
  \department{School of Electronic Engineering and Computer Science}
  \city{London}
  \country{UK}
}

\begin{document}
%Over the past five years, a growing body of qualitative research has examined these early transformations across diverse roles, studios, and contexts. However, the current moment calls for a broader, more integrated overview that situates these emerging findings within wider game production trajectories.

\begin{abstract}
\ac{GenAI} is currently reshaping game development practices, production pipelines, and value networks in an unprecedentedly pervasive manner with cascading consequences remaining unclear. In the last five years since \ac{GenAI}'s inception, a growing body of qualitative research has explored these early transformations from different settings and demographic angles. However, these studies often contextualise and consolidate their findings weakly with related work; for research to keep up with and support stakeholders in this development, the current moment calls for a synthesis of the findings emerged thus far. Here, we adress this need through a qualitative research synthesis via meta-ethnography. We followed PRISMA-S to systematically search the relevant literature from 2020-2025, including major HCI and games research databases. We then synthesised the ten eligible studies, conducting reciprocal translation and line-of-argument synthesis guided by eMERGe, informed by CASP quality appraisal. We identified nine overarching themes, provide recommendations, and contextualise our insights in wider game production trajectories. With this work, we seek to provide practitioners, researchers and policy-makers with grounded insights to guide practice, research and governance.
\end{abstract}


\maketitle

\section{Introduction}
\label{sec:introduction}
% eMERGe criteria 1,2,3 ( rationale & context + study aim and purpose)

\acf{GenAI} has disrupted and is transforming creative workflows across most industries and forms of creative practice. Game development, traditionally an early adopter of new technologies, is no exception. Transformation through \ac{GenAI} has been identified as a major and lasting trend in this domain \citep{kultima2025gamingtrends}. The rapid uptake of \ac{GenAI} in commercial game production has also been confirmed through publishing data: recent reporting on \emph{Steam} submissions indicates that at least\footnote{This likely under-represents the actual scale of adoption since disclosure is inconsistent. Reflecting this trend, \emph{Valve} (a video games company) has updated its \emph{Steam} (online platform for buying, downloading, and playing video games) content survey to require disclosure of both pre-generated AI content (produced during development) and live-generated AI content (produced during gameplay).} 7\% of all games released on the platform now disclose some form of AI usage for content creation, up from only 1\% the year before \citep{totallyhuman2025steam}. Around 60\% of disclosed implementations involve visual asset generation, demonstrating how art pipelines are a central entry point for GenAI in game production. 

The adoption of \ac{GenAI} in game production calls for researchers not only to improve systems and interaction modalities, but also to shape a societally and economically sustainable future of human-AI co-creation \citep{kantosalo2020five} and support the advancement of games as diverse cultural artefacts. Due to its complex and rapidly evolving nature, we deem \emph{qualitative research} the prime vehicle to explore this phenomenon. Existing qualitative studies have, amongst others, investigated how \ac{GenAI} reshapes workflows, raises ethical questions, and transforms relations between developers, artists, and players. Crucially though, while scholars have dedicated studies to the adoption of \ac{GenAI} in game production at a rarely seen and still increasing level of intensity, these investigations are often \emph{limited} to specific types of e.g.~target demographic and production setting, or only a narrow aspect of the overall phenomenon. In addition, the research landscape remains \emph{fragmented}, reflecting the \enquote{one-off problem of failing to build upon prior work} \citep{drisko2020qualitative} -- a broader pattern in qualitative research. While some qualitative studies (Sec.~\ref{sec:related-consolidation}) notably relate to previous work more extensively, this is (naturally) done in the service of distinguishing and motivating their own focus or contributions, and not from a neutral perspective. This makes it difficult for researchers, practitioners, and policymakers to grasp the big picture and assess how strongly smaller observed phenomena are empirically supported. A \emph{synthesis} is needed to translate insights across studies, identify consistencies and contradictions, and highlight areas requiring further exploration -- both in terms of the studies phenomena and the methodology used.

Against this backdrop, our aim was to \emph{systematically review and synthesise qualitative research on the adoption and impact of \ac{GenAI} in game production}. Specifically, we seek to answer the following research questions (RQs):
\begin{enumerate}
%AD: Rephrased RQs based on R3's comments
\item How do existing qualitative studies differ, e.g.~in research questions, methodology, demographics and setting?
\item How do the findings from these qualitative studies characterise the ways in which \ac{GenAI} is adopted, used and perceived within game development workflows?
\item In what ways do findings across studies converge, complement, or contradict one another?
\item Which conceptual, empirical, or methodological gaps remain? How might these shape future research priorities?
\end{enumerate}

We address these questions through a \emph{meta-ethnography} as the arguably most established, interpretive \ac{QRS} method (see Sec.~\ref{sec:methods} for further justification). \ac{QRS} has been highlighted as a systematic and reasonably robust method \citep[p.~69]{campbell_evaluating_2011} capable of identifying saturated as well as underexplored areas of research \citep[p.~120]{campbell_evaluating_2011}. To ensure methodological rigour, we follow the \emph{eMERGe} reporting guidelines \citep{france_emerge_2019} for meta-ethnography, documenting our search and screening via \emph{PRISMA-S} \citep{rethlefsen_prisma-s_2021}, and appraising primary study quality with the \emph{\acf{CASP}} \citep{long_optimising_2020}. We focus on studies from 2020-2025 investigating the use of publicly available \ac{GenAI} in \enquote{offline} game production (see Sec.~\ref{sec:selection-eligibility} for further scoping). Drawing together insights from primary research, this synthesis not only consolidates a scattered evidence base, but also probes the generalisability of observed phenomena.

The result is the first integrated qualitative evidence base on \ac{GenAI}'s impact on game production, offering \emph{industry stakeholders} a clearer picture of emerging practices to adopt or avoid while equipping \emph{researchers}, \emph{funders} and \emph{policymakers} with grounded insights to guide future investigations, support decisions and governance. More specifically, we make at least three \emph{contributions}. Firstly, in answering \textbf{RQ1}, we map the existing qualitative research landscape on \ac{GenAI} in game development in terms of what studies sought out to do, in which context, and by which means, thus providing a much needed overview and signpost to studies with specific foci. Secondly, through interpretative work on \textbf{RQ2/3}, we abstract the core phenomena found by qualitative researchers and make transparent how they are similarly supported, enriched or contradicted by different studies, thus providing a nuanced overview of a rapidly evolving research complex to guide practice, research and governance. This includes, for instance, how practitioners negotiate the integration of GenAI in their creative and technical workflows, and how \acs{GenAI} reshapes developers' perceptions of creative agency, authorship, and craft accountability within game development. Based on these insights, we derive implications and recommendations for all stakeholders. Thirdly, via \textbf{RQ4} and supported by \textbf{RQ1/3}, we provide the necessary insights to inform future research priorities or caution decision-making based on incomplete insights. 

\section{Background}
\label{sec:background}
% eMERGe criteria 1 ( context specifically)

While generative systems date back to early computer art, e.g., \citep{mccorduck1991aaron,Sims91}, today's \acf{GenAI} refers mainly to machine learning systems, particularly Transformers \citep{vaswani2017attention} and Diffusion Models \citep{ho2020denoising}. \ac{GenAI} is reshaping game design and development, creating both opportunities and challenges across creative and technical processes \citep{ratican2024adaptive}. We briefly outline the advancements of \ac{GenAI} relevant to this work, focusing on text, image, 3D and audio content.

The development of \acp{LLM} has transformed natural language processing, enabling models to generate highly realistic text from minimal examples \cite{brown2020language}. GPT-1 (2018) \citep{radford2018improving} showed that strong performance across diverse language tasks was possible, paving the way for today’s state-of-the-art systems.
\acp{LLM} continue to evolve rapidly, shaping both game writing and technical workflows. GPT-3 and 3.5 \citep{brown2020language} popularised conversational interaction through \emph{ChatGPT}, while GitHub \emph{Copilot} demonstrated how code generation could accelerate engine scripting and gameplay logic. Google's latest \emph{Gemini 2.5} model adds advanced multimodal reasoning, allowing integration of text, image, and audio in design workflows \citep{oppenlaender2022creativity, inie2023}. Unity \emph{Muse} \cite{unity_muse} (introduced in Unity 6) now embeds LLM assistance directly into engine editor workflows, enabling code completion, asset search, and interaction prototyping out-of-the-box.

With advances in \acp{LLM}, \ac{TTIG} systems emerged \citep{brown2020language}. Caption-conditioned models had already shown that text could guide image generation \citep{mansimov2015generating}, and CLIP \citep{radford2021learning} strengthened the field by learning visual concepts from natural language. OpenAI's \emph{DALL·E} (2021) \citep{openai_dalle, ramesh2021zero} combined a Transformer \citep{vaswani2017attention} with CLIP \citep{ramesh2022hierarchical}, while \emph{DALL·E 2} \citep{openai_dalle2} replaced the Transformer with Diffusion \citep{ho2020denoising}, producing images that were more realistic and faithful to prompts. Systems such as \emph{DALL·E 3}, \emph{Midjourney} \citep{midjourney}, \emph{Stable Diffusion} \citep{stable_diffusion, rombach2022high} are now widely used for creative work. 

Naturally, AI's capabilities have been extended from two- to three-dimensional content generation. Text2Mesh \citep{michel2022text2mesh} introduced text-driven colour and geometry editing of meshes, while CLIP-Forge \citep{sanghi2022clip} proposed zero-shot text-to-3D generation using CLIP embeddings without paired data. Recent systems can now transform simple text descriptions into 3D models \citep{dai2023towards}. Tools such as \emph{Alpha 3D} \cite{alpha3d}, \emph{Luma AI} \cite{luma_ai} and \emph{Meshy.ai} \cite{meshy_ai} create textured, game-ready assets from text or image prompts. \emph{Promethean AI} \cite{promethean_ai} assists designers by automating environment assembly \citep{begemann_gaidev_2024}, while \emph{Genie 3} generates interactive 3D worlds from a single prompt with evolving dynamics \citep{parker-holder_fruchter_2025_genie3}. Similarly, Tencent’s \emph{Hunyuan-Game model} \cite{li2025hunyuan} extends generation to entire asset sets, including characters, effects, and video snippets.% trained on billions of game-specific resources.

As our final group of \ac{GenAI} to introduce, generative audio systems are beginning to automatise sound and voice production. Tools such as \emph{Soundraw} \cite{soundraw} provide AI-assisted adaptive music composition, while \emph{ElevenLabs} \cite{elevenlabs} enables lifelike voice acting with controllable accents and emotions. Diffusion-based systems like \emph{Producer.ai} \cite{producer_ai} and research frameworks such as \emph{Meta's Audiocraft} \cite{copet2023audiocraft} expand possibilities for procedural soundscapes. These developments hint at future workflows where background scores, diegetic sound effects, and even NPC voices can be prototyped without traditional recording pipelines, providing opportunities but also raising concerns about replacing skilled people \citep{elgammal2020artists, wingstrom2022redefining}.

\section{Related Work}
To our knowledge, no prior work has systematically synthesised qualitative research on the impact of \ac{GenAI} on game development or developer experiences. We position our synthesis relative to adjacent reviews within games, clarifying differences in scope and methods. We also relate to the few primary studies captured by our synthesis that connect their findings with previous work, complimenting their scholarship and further motivating our approach.

\subsection{Reviews on Generative AI in Games}
\citet{sweetser2024large} conducted a scoping review and reflexive thematic analysis of early research (2022-Spring 2024) on LLMs and video games across five application themes. While most reviewed papers are technical, some qualitative observations from developers' interaction with \ac{GenAI} are included, making it closest to our work. Key differences lie in technical focus, methodology, and timeframe. We focus exclusively on qualitative studies of developer experiences, over a longer timeframe (2020-Summer 2025), and, through meta-ethnography, provide a more systematic and conceptually richer synthesis of existing work, which preserves the interpretations of the primary literature. 

Similarly, \citet{moon2025generative} combine qualitative, empirical and quantitative technical work to assess the relevance and challenges of \ac{GenAI} in educational game design across mutiple stakeholder groups. The review and analysis methodology as well as exact timeframe of covered literature is left opaque, and the report does not connect claims directly to specific qualitative insights from the primary literature. We, in contrast, provide a systematic, grounded synthesis of exclusively qualitative studies on \ac{GenAI}'s impact on game production, focusing on developers' insights while covering diverse production contexts and game types, including educational games.

Other reviews draw on technical literature even more substantially to identify opportunities and challenges in using specific \ac{GenAI} technologies -- often research prototypes -- for game development. 
Focusing on \ac{GPT} models and technical venues, \citet{yang2024gptgames, yang2025gptGames} provide a scoping review of GPT applications to games from 2020 to 2024 across \ac{PCG}, mixed-initiative game design, mixed-initiative gameplay, playing games, and game user research. The reviews do not report a specific literature search and analysis standard. Other technical reviews embrace a broader variety of AI but narrow their focus on specific applications in the game development pipeline. Using PRISMA, \citet{wu2025genAICharacter} systematically review papers demonstrating and advocating the use of \ac{GenAI} for game character creation across concept, modelling, animation, and behaviour design (2019-Autumn 2024). While their emphasis is on design-time support, \citet{maleki2024procedural} review methods to implement \ac{PCG} primarily at runtime, with a focus on \acp{LLM}. Their review and GPT-3.5d analysis covers literature from 2019 to 2023 but does not mention a specific standard.  \citet{ribeiro2024imageGenAI} employ PRISMA to systematically review applications of image generation models for image-based game asset production, with a focus on image quality metrics. With papers from 2016 to 2023, the review also covers techniques that are not associated with \ac{GenAI} in the popular sense (cf.~Sec.~\ref{sec:background}).\\ 

In contrast to these reviews, we focus exclusively on \emph{synthesising} rich, \emph{qualitative} insights from studies dedicated to documenting \emph{game developers' experience} with \ac{GenAI}. Similar to the technical reviews mentioned above, this data also conveys opportunities and shortcomings of \ac{GenAI} applications but is grounded in \emph{concrete, in-situ insights from actual game development} rather than researcher proposals on prototype potentiality. In common with these reviews, we also seeks to identify gaps and opportunities in research. However, these concern \emph{gaps in evidence and methodology}, not in technology or specific applications. To support the transfer of insights, we focus on serviced, publicly available systems and exclude work on non-public research prototypes. Moreover, we focus on \ac{GenAI} technologies more generally, rather than specific flavours, and on its impact on game development across contexts, rather than a specific aspect of the pipeline. Arguably most importantly, we conduct a methodologically rigorous, purely qualitative research synthesis via meta-ethnography \citep{noblit_meta-ethnography_1988,cahill_metaethnography_2018}, following established standards across the steps of literature search \citep[PRISMA-S,][]{rethlefsen_prisma-s_2021}, quality appraisal \citep[CASP,][]{long_optimising_2020} as well as analysis and synthesis \citep[eMERGe,][]{france_emerge_2019}, facilitating the emergence of new conceptual insights. Epistemologically speaking, while the existing reviews primarily aggregate first-order interpretations from researchers, we interpret researchers' second-order constructs of their participants' experiences into third-order constructs as part of our synthesis, using participants' first-order interpretations to ground the analysis and synthesis (see Sec.~\ref{sec:methods_qrs} for a distinction between aggregative/interpretative synthesis and1\textsuperscript{st}/2\textsuperscript{nd}/3\textsuperscript{rd}-order interpretations). 

\subsection{Consolidation in Primary, Empirical Studies of GenAI in Games}
\label{sec:related-consolidation}
Across our corpus of primary studies, only six out of 10 relate their findings to existing empirical work within game development. Here, the depth and scope of  integration vary considerably. The first studies on the topic \citep[e.g.][]{grow_chatgpt_2023,vimpari_adapt_or_die_2023} naturally integrate their findings mostly with work outside games. However, even among later studies, substantive cross-reference to other empirical studies in game production remains rare. The most notable exceptions are \citet{boucher_resistance_2024}, who situate their findings within debates on authorship and labour in- and outside games, and \citet{alharthi_genaicreativity_2025}, who primarily relates to empirical studies to link production practices with broader questions of creativity and industry adoption.\\

We provide a more comprehensive, neutral and systematic re-integration of these individual findings with the larger and up-to-date body of qualitative research. Given the interpretative nature of meta-ethnography, the partial integrations in the primary literature above can be used as contrast and complement our synthesis.  

\section{Methods}
\label{sec:methods}
This paper aims to draw a bigger picture from existing qualitative research on how \ac{GenAI} has been adopted, received, and integrated within game development. To this end, we conducted a \emph{\acf{QRS}} and a quality appraisal of the underlying studies. While common in e.g.~medical research \cite{cahill_metaethnography_2018}, our methods are only gaining traction in HCI and games research. Here, we introduce and motivate the core methods of this study. We sketch our collaboration across all methods in Sec.~\ref{sec:limitations-reflexivity}, and provide a detailed account, reflexive procedures and positionality in Appx.~\ref{app:reflexivity}.

\subsection{Qualitative Research Synthesis}
\label{sec:methods_qrs}
\acf{QRS} encompasses methods for systematically combining the data or interpretive findings from multiple qualitative studies to
generate new knowledge and theory \cite{drisko2020qualitative} about a phenomenon. \emph{aggregative \ac{QRS} methods} typically take a realist/pragmatist epistemology to pool and describe the original data (e.g., interviews) from large amounts of primary studies. In contrast, \emph{interpretive methods} take an intepretivist/constructivist stance to synthesise the interpretive findings (e.g., codes developed on interviews) from often smaller sets of primary studies \cite{drisko2020qualitative}.

We conducted an \emph{interpretive \ac{QRS}} because (i) it can extend prior conceptualisation and theory; (ii) comprehensive raw data are not available in most relevant primary studies (Sec.~\ref{sec:results_included_studies}); (iii) it enables critical appraisal of the strengths and weaknesses of the original research; and (iv) it makes contextual and human diversity features more apparent across studies \citep[][quoting Paterson, 2012]{drisko2020qualitative} -- \citet[p.~123]{campbell_evaluating_2011} note its empowering function in involving \enquote{methods for combining multiple voices to seek new interpretations, rather than dismissing single case studies as locally bound}. While most relevant to our second RQ (Sec.~\ref{sec:introduction}), (i)-(iii) are pivotal to addressing our third and fourth RQs . 

Following common \ac{QRS} terminology \citep[e.g.][]{noblit_meta-ethnography_1988, britten2002using}, we refer to raw data in the primary literature as \enquote{1\textsuperscript{st}-order interpretations}, emphasising that e.g.,~interview statements also constitute an individual's interpretation of their experience. \enquote{2\textsuperscript{nd}-order interpretations} then denote authors' interpretations of this data through qualitative analysis, shaped by theories and individual perspectives. Finally, \enquote{3\textsuperscript{rd}-order interpretations} emerge from the synthesis approach by critically comparing these second-order interpretations. We later extend this taxonomy slightly in Sec.~\ref{sec:execution-extraction}.

\subsection{Meta-Ethnography}
\label{sec:methods-meta-ethnography}
% eMERGe Items 4 (rationale, expanded from Introduction) - Still needs supporting literature for rationale 
We conducted a meta-ethnography, the pioneering and most widely used interpretive \ac{QRS} method \citep{drisko2020qualitative}. Developed by \citet{noblit_meta-ethnography_1988} and later extended \citep{france_emerge_2019}, it introduces seven steps widely adopted in interpretive \ac{QRS}: 

\begin{multicols}{2}
\begin{enumerate}
\item Selecting meta-ethnography and getting started
\item Deciding what is relevant
\item Reading included studies
\item Determining how studies are related
\item Translating studies into one another
\item Synthesising translations
\item Expressing the synthesis
\end{enumerate}
\end{multicols}

Meta-ethnography enables conceptual translation across studies while preserving their contextual and epistemic integrity \citep{drisko2020qualitative}. In step (5), we specifically employ \emph{reciprocal translation}, relating 2\textsuperscript{nd}-order interpretations across studies while noting similarities and differences \citep{drisko2020qualitative} and supporting with 1\textsuperscript{st}-order interpretations when needed.

The \emph{synthesis} (step 6) is an inductive, iterative process in which the researcher produces 3\textsuperscript{rd}-order interpretations as conceptual abstractions by (potentially) re-analysing, relating and extending the original study authors' 2\textsuperscript{nd}-order interpretations, while carefully preserving their grounding in the 1\textsuperscript{st}-order interpretations. Although described as separate phases, synthesis can start during translation -- as was also the case here. The synthesis culminates in a \emph{line-of-argument} narrative in which 2\textsuperscript{nd}-order interpretations from different primary studies are re-narrated, structured by our new 3\textsuperscript{rd}-order interpretations. The result \enquote{says something about the whole [phenomenon] based on studies of the parts} \citep[p.~64]{campbell_evaluating_2011} and forms, also due to its interpretive nature, a new complement to the primary studies \citep[p.~121]{campbell_evaluating_2011}.

For precision and transparency, we report our meta-ethnography in accordance with the 19 criteria outlined in the dedicated eMERGe framework \cite{france_emerge_2019}.

\subsection{Qualitative Research Appraisal}
\label{sec:methods-appraisal}
Acknowledging the ongoing debate on the value of quality appraisal in qualitative synthesis \cite{long_optimising_2020,cahill_metaethnography_2018, dixon-woods2008}, we incorporate formal study quality assessment. This can also contribute as interpretive aid and encourage the closer and repeated reading of primary studies \citep[p.~122]{campbell_evaluating_2011}. Following \citeauthor{majid_appraising_2018}'s \citeyearpar{majid_appraising_2018} decision tool, we selected the Critical Appraisal Skills Programme (\ac{CASP}) Qualitative Checklist as (1) the most common appraisal tool in \ac{QRS} that is (2) domain-agnostic, (3) short, and (4) relatively easy to employ. We use \citeauthor{long_optimising_2020}'s \citep{long_optimising_2020} updated and optimised version. 

In line with guidance that appraisal should inform, but not dictate, interpretive decisions in meta-ethnography \citep{campbell_evaluating_2011, cahill_metaethnography_2018}, we use quality appraisal not to determine study inclusion, but to (1) support interpretive confidence of each study's findings during extraction and translation \cite{long_optimising_2020} and to (2) critically reflect on methodology as a lens to pinpoint gaps in research (our fourth RQ, Sec.~\ref{sec:introduction}). We report this weighting alongside the references to the primary sources in the translation maps and the code book to make clear where interpretive grounding is stronger and where further work is needed.

To reconcile a study's conceptual contribution with its methodological rigour, we decided not to adopt the popular taxonomy of 'Key Paper', 'Satisfactory', and 'Fatally Flawed' \citep{dixon-woods2008}. For one, the latter category might appear antagonising and hinder our goal to constructively encourage methodological rigour. Moreover, 
\citet{campbell_evaluating_2011} found that a study's methodological quality (e.g.~\enquote{fatally flawed}) was not always reflective of its conceptual richness (e.g.~a \enquote {key paper}); similarly, \citet{dixon-woods2008} document reviewers' dilemmas when assessing studies they found to be highly relevant despite flaws in their research conduct. 
Therefore, our approach separates these two judgements. Following the updated \ac{CASP} framework \citep{long_optimising_2020}, each paper is assigned a quality rating of Low, Medium, or High. Complementing this, we flag studies that function as \enquote{Key Papers} conceptually, i.e.~ that are \enquote{conceptually rich and could potentially make an important contribution to the synthesis} \citep{malpass_medication_2009}. Our two-pronged appraisal system ensures that we are not overlooking conceptual innovation whilst still maintaining a clear view of the methodological rigour underpinning our synthesis. 

\section{Selection of Primary Studies}
\label{sec:study_selection}
Searching and screening literature for a qualitative synthesis is acknowledged to be a challenging and multi-faceted process \cite{cahill_metaethnography_2018}. 
%eMERGe Item 5 - search strategy rationale. 
To foster transparency and reproducibility, we implement the \ac{PRISMA-S} guidelines for reporting literature searches in systematic reviews \citep{rethlefsen_prisma-s_2021}, thus adding detail to \acs{eMERGe} steps 5--7. The PRISMA process is summarised in Figure ~\ref{fig:prisma}.

\subsection{Eligibility Criteria} 
% eMERGe item 5 - Search Strategy Rationale %
\label{sec:selection-eligibility}

\paragraph{Inclusion}
Our eligibility criteria (Table ~\ref{tab:eligibility}) were developed iteratively and collaboratively (details in Appx.~\ref{app:reflexivity}) at the outset of the review. We included studies that report \emph{original empirical research} using \emph{qualitative data collection} and \emph{qualitative analysis} methods to examine the use of \emph{GenAI} systems within \emph{game development} contexts. We deemed mixed-methods studies eligible if the qualitative component was substantial and clearly reported. % eMERGe item 5 - Search Strategy Rationale %
With this, we express our focus on studies that offer insight into GenAI's practical impact over speculative or theoretical perspectives. % eMERGe item 5 - Search Strategy Rationale + PRISMA 9 - Limits and Restrictions %
To ensure relevance to current practices and a match of what is dominantly discussed under the umbrella of \enquote{\acl{GenAI}}, only studies published in the \emph{last five years} (January 2020 - June 2025) were included, i.e.~since publicly accessible GenAI tools became widely available to our demographic. Studies were considered if they reported on participants' \emph{direct experience of game development}, either in professional roles or through situated creative production such as game jams or study assignments. We deliberately chose this wider focus as it includes people outside game industry who may eventually join the industry or contribute to game production from outside.

\begin{table}[h!]
\caption{Eligibility Criteria}
\Description{Table summarising inclusion criteria for the review. Lists publication period (Jan 2020–June 2025), study design (qualitative or mixed-methods empirical work), study focus (application of GenAI in game development), participant requirements (active in development), and scope (publicly available GenAI).}
\label{tab:eligibility}
\begin{tabularx}{\textwidth}{l l}
\toprule
\textbf{Criterion} & \textbf{Description} \\
\midrule
\addlinespace
Publication Period & January 2020 -- June 2025. \\
\addlinespace
Study Design & Reports original empirical research using qualitative or substantial mixed-methods designs. \\
\addlinespace
Study Focus & Examines the application and experience of GenAI in game development contexts.  \\
\addlinespace
Participants & Must act in a game development capacity (e.g., professionals, game jam participants). \\
\addlinespace
Scope of Technology & Publicly available GenAI systems.  \\
\bottomrule
\end{tabularx}
\end{table}

% eMERGe Item 7: Selecting primary studies
% PRISMA-S Item 9: Limits and restrictions
\paragraph{Exclusion}
We excluded work that relied on custom-built or non-public \ac{GenAI} systems, as the corresponding findings may not easily generalise. We moreover excluded work that focused on technical implementation and evaluation, as well as studies that gathered data from players, students, or other stakeholders outside of a development capacity. 

% eMERGe item 5 - Search Strategy Rationale %
% PRISMA-S Item 9: Limits and restrictions
\paragraph{Limits and restrictions}
We limit our synthesis to qualitative studies in acknowledgement that \ac{GenAI} is a relatively new and rapidly evolving phenomenon within the game development industry, and thus benefits from the rich, nuanced, and exploratory insights offered by qualitative inquiry. We acknowledge the value of quantitative inquiry, which can be further informed through the qualitative insights and research gaps identified here.

\subsection{Search Strategy}
\label{sec:search_strategy}
% eMERGe item 6 - Search Process %
We conducted a systematic literature search across academic databases relevant to HCI and games research, complemented with manual and open searches. % PRISMA-S Item 1: Database name
We initially searched the ACM Digital Library and Scopus via Elsevier, as they index the major publication venues in HCI and technical game research. 
% PRISMA-S Item 4: Online resources and Browse
We further conducted manual searches of the DiGRA and the Foundations of Digital Games (FDG) proceedings to account for relevant work that may not be well indexed elsewhere. Finally, we searched openly on Google Scholar to identify preprints and peer-reviewed work published in other venues. Two additional measures taken to expand the identified corpus are described in Sec.~\ref{sec:study_selection_selection_process}.

% eMERGe Item 6: Search processes
% PRISMA-S Item 8: Full search strategies
The search terms were developed iteratively, informed by prior literature and exploratory queries. Terms were grouped into four primary blocks: (1) game development contexts, (2) \ac{GenAI} systems and architectures, (3) qualitative research design, and (4) conceptual framings and interaction modalities. Searches were piloted in the ACM Digital Library and Scopus and then extended to the conference proceedings and Google Scholar. Boolean operators and wildcards were used to manage linguistic variation, and search strings were adapted to suit the syntax requirements of each platform. Full search strategies, including all query variants, are included as Supplementary Materials.

\begin{figure}[htbp]
    \centering
    \includegraphics[width=0.8\textwidth]{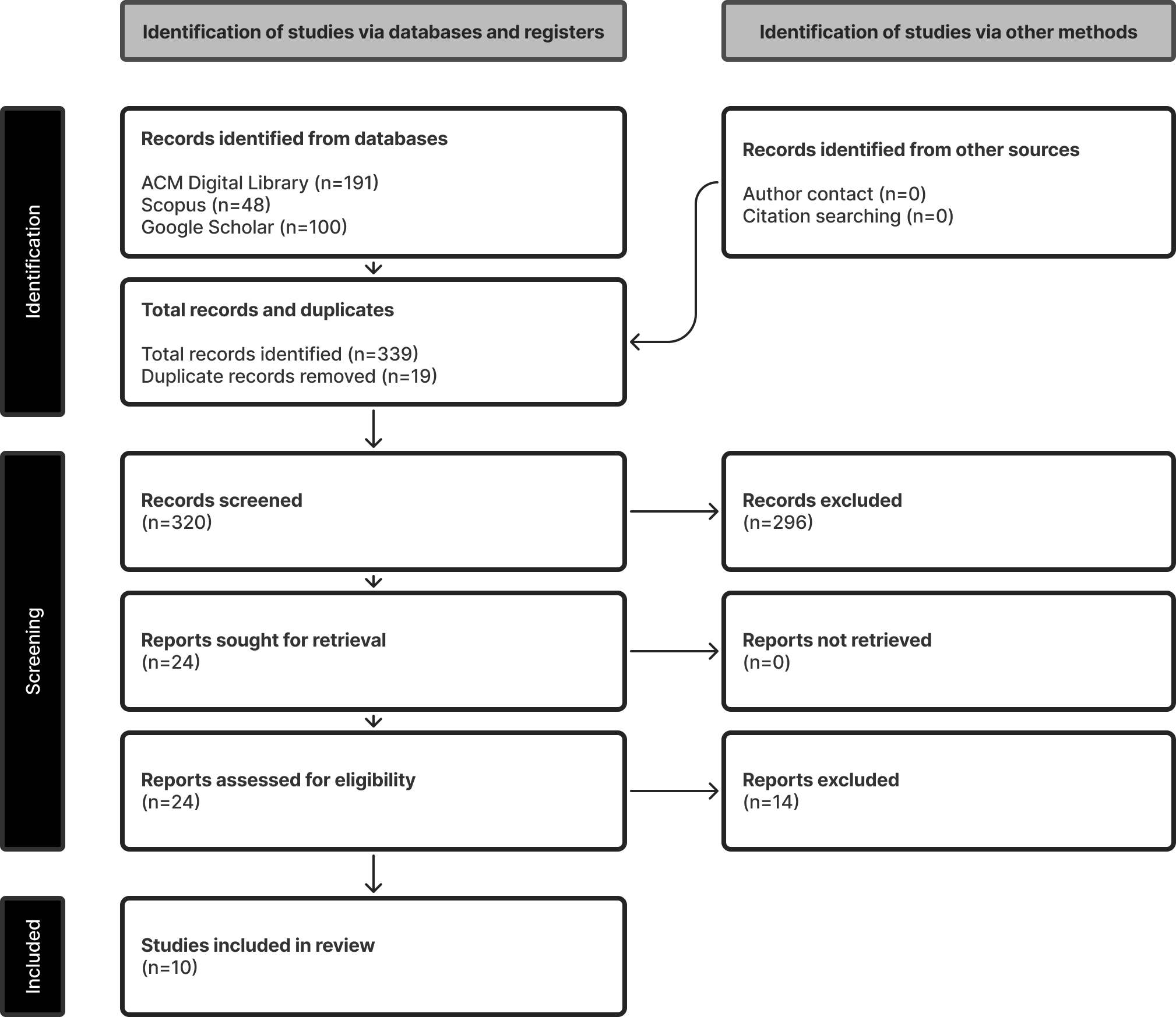}
    \caption{PRISMA Flow Diagram}
    \Description{A PRISMA-style flowchart summarising study selection. From 339 records initially identified, 19 duplicates were removed, 320 were screened, and 24 reports were retrieved. After eligibility assessment, 10 studies were included in the final review. The flow diagram uses boxes and arrows to show this stepwise filtering process.}
    \label{fig:prisma}
\end{figure}

\subsection{Selection Process}
\label{sec:study_selection_selection_process}
% eMERGe item 7 - Selecting Primary Studies + PRISMA S items 9, 11 - fulfilled %
Individual database yields are reported in Fig.~\ref{fig:prisma}. For Google Scholar, the first 100 results were sorted by relevance and selected for review. After manually removing duplicates, 320 records remained. Screening and selection followed a consensus-based approach (details in Appx.~\ref{app:reflexivity}), ensuring agreement on inclusion decisions at both abstract/title screening and full-text review stages. Based on the review of titles and abstracts, 
we identified seven clearly eligible and 17 borderline studies that were subjected to detailed evaluation via full-text close reading with a focus on methodology and findings. This closer inspection confirmed three more papers, establishing a preliminary corpus of 10 studies. While modest in scale, the final corpus reflects the specificity of our eligibility criteria as encoded in the search query.

To ensure our primary studies pool was as comprehensive and up-to-date as possible, we took two further steps. First, we conducted backward citation searching, manually examining the bibliographies of the ten included articles for any relevant studies. Second, we contacted the first and last authors of these publications to enquire about any in-press or forthcoming work meeting our eligibility criteria. Neither step yielded further studies; while we had constructive dialogue with authors, their current research was either out of scope or still in progress when our analysis began. 

\section{Quality Appraisal}
\label{sec:execution-qa}
The qualitative research appraisal (Sec.~\ref{sec:methods-appraisal}) of each paper was done collaboratively (details in Appx.~\ref{app:reflexivity})  with the modified 11-question CASP checklist \cite{long_optimising_2020}, and detailed justifications are provided as Supplementary Materials. Following procedure by \citet{long_optimising_2020}, we selected tipping-point criteria to decide on borderline weightings, namely rigour of the analysis and the clarity with which findings were grounded in the data. Tbl.~\ref{tab:casp-summary} provides a per-criterion overview, highlighting that all studies clearly stated the aims of research (criterion 1) and proposed an appropriate research design (criterion 3). We found the highest variation for criteria 7--9, with several studies not or only somewhat considering the researcher--participant relationship, addressing ethical issues, or performing a rigorous data analysis. The resulting appraisal weightings (\enquote{Low},\enquote{Medium},\enquote{High}) for individual papers are presented in Tbl.~\ref{tab:dataQA}. 

\begin{table}[ht]
\centering
\caption{Summary of CASP appraisal of included studies (N = 10). Note: CASP scores reflect independent criteria, i.e. a study may receive a \enquote{Yes} for design appropriateness even if its later execution (e.g., data collection or analysis) was flawed.
}
\Description{Quality appraisal table showing responses to 11 CASP criteria across 10 studies. Most criteria were rated ``Yes'' (e.g., aims, design, value), while some showed weaker evidence (e.g., researcher–participant relationship and recruitment strategy). Few ``No'' or ``Can't tell'' responses.}
\begin{tabular}{p{9.3cm}cccc}
\toprule
\textbf{CASP Criterion} & \textbf{Yes} & \textbf{Somewhat} & \textbf{No} & \textbf{Can't Tell} \\
\midrule
1. Was there a clear statement of the aims of the research?                & 10 & 0 & 0 & 0 \\
2. Is a qualitative methodology appropriate?                               & 7  & 3 & 0 & 0 \\
3. Was the research design appropriate to address the aims?               & 10 & 0 & 0 & 0 \\
4. Are the study’s theoretical underpinnings clear and coherent?          & 9  & 1 & 0 & 0 \\
5. Was the recruitment strategy appropriate to the aims?                  & 4  & 1 & 0 & 5 \\
6. Was the data collected in a way that addressed the research issue?     & 8  & 2 & 0 & 0 \\
7. Was the researcher–participant relationship adequately considered?     & 2  & 1 & 4 & 3 \\
8. Have ethical issues been taken into consideration?                     & 5  & 2 & 1 & 2 \\
9. Was the data analysis sufficiently rigorous?                           & 4  & 3 & 3 & 0 \\
10. Is there a clear statement of findings?                               & 9 & 1 & 0 & 0 \\
11. How valuable is the research?                                         & 10 & 0 & 0 & 0 \\
\bottomrule
\end{tabular}
\label{tab:casp-summary}
\end{table}
\section{Data Extraction}
\label{sec:execution-extraction}
%eMERGe item 9 - data extraction & reading approach%

Each included study was initially subjected to a close reading by one team member to extract the synthesis materials (see Appx.~\ref{app:reflexivity} for collaboration details). Because \enquote{all [2\textsuperscript{nd}-order] interpretations must be grounded in the [primary] texts to be synthesised} \citep{noblit_meta-ethnography_1988}, we extracted 2\textsuperscript{nd}-order interpretations if their grounding in 1\textsuperscript{st}-order interpretations was made explicit, and were inclusive where the link was implicit but clear in context. CASP ratings (Sec.~\ref{sec:methods-appraisal}) informed this judgement pragmatically: in studies appraised as higher quality, implicit grounding was more readily accepted as sufficiently evidenced, whereas in lower-rated studies we sought explicit textual support before extracting an interpretation. Extraction and subsequent translation was done in Atlas.ti 25 (Desktop), treating each primary source as separate document.

We encountered two additional challenges. Firstly, primary studies feature interpretations of similar phenomena but at varying depth (e.g. one code or a theme with subcodes). Secondly, 2\textsuperscript{nd}-order interpretations are given at various levels of conciseness (e.g.~a code vs. an interpretive paragraph) and/or scarcity (e.g.,~some authors letting participants' 1\textsuperscript{st}-order interpretations speak largely for themselves). To address the risk of rich, readily extractable interpretations biasing the extraction of less concise, scarcer or absent 2\textsuperscript{nd}-order interpretations, we first engaged with studies featuring the latter. To systematically handle conceptual depth differences, we adopted a two-step approach to the extraction:
\begin{enumerate}
    \item An initial extraction of top-level 2\textsuperscript{nd}-order interpretations was conducted across all studies.
    \item We then returned to studies where an already identified 2\textsuperscript{nd}-order interpretation was elaborated     (e.g., \citet{boucher_resistance_2024} on ethical resistance) to extract their more granular sub-themes in explicit relation to the parent concept. 
\end{enumerate}  Our preservation of context and meaning of interpretations within and across studies followed guidance by \citet{france_emerge_2019}: we  retained the original wording and framing of 2\textsuperscript{nd}-order interpretations, documented their provenance, and preserved the hierarchical links between interpretations provided by the primary studies. Where papers offered extensive 1\textsuperscript{st}--order but minimal researcher interpretation\citep[e.g.][]{grow_chatgpt_2023}, we conducted our own thematic analysis to generate what we term and later report as \enquote{2\textsuperscript{nd}-order interpretations augmented}. We clearly highlight these new interpretation in the visual documentation of our reciprocal translation, also showing all extracted interpretations (Sec.~\ref{sec:translation}). Extracted interpretations were reviewed collaboratively, leading to refinement or merging to enhance clarity and coherence prior to synthesis. We engaged in memo-taking to preserve these insights for the analysis to follow.

\begin{figure}[t!]
    \includegraphics[width=1\textwidth]{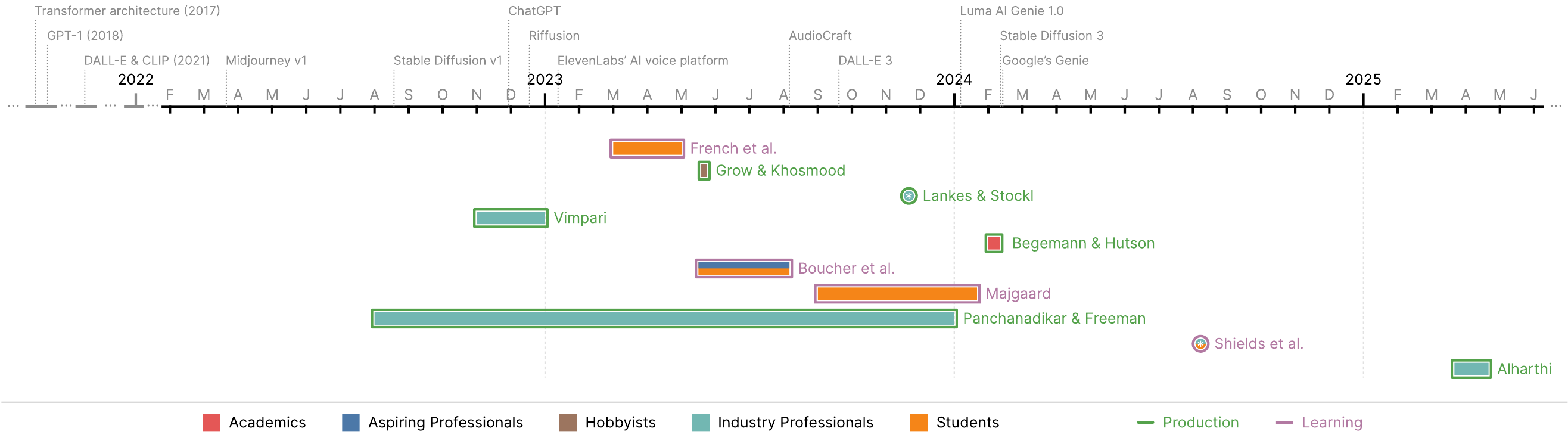}
    \caption{Timeline with key \ac{GenAI} release dates and data collection intervals for the ten included studies. Studies are colour-coded for demographic focus, and outlines distinguish the study context ({\color[HTML]{54A24B}green} for production; {\color[HTML]{B279A2}purple} for learning). Detailed data for \citep{alharthi_genaicreativity_2025} and \citet{begemann_gaidev_2024} were obtained via direct correspondence. Data for \citet{boucher_resistance_2024} and \citet{majgaard_pilot_2024} were supplemented with information from public online sources. For \citet{panchanadikar_solodev_2024}, the interval reflects the date range of collected social media posts. It was not possible to determine the data collection period for \citet{lankes_ai-powered_2023} and \citet{shields_generatingtogether_2024} and these are marked in a circular shape.}
    \Description{A timeline diagram showing the time interval of the data collection in each study. The timeline also shows key release dates from milestone GenAI systems. The figure illustrates how the majority of the studies took place between 2023 and 2024 but are scattered within with typically short data collection intervals.}
    \label{fig:timeline}
\end{figure}

\begin{table}
\caption{Overview of the included studies (in chronological order) as well as their research questions or objectives. Quotation marks denote author-stated content; unquoted items are inferred from the paper. Further details are provided in subsequent tables.
}
\Description{Table listing 10 included studies with citation, title, and research questions or objectives. Objectives include exploring GenAI in education, professional practice, indie development, and game jams, with emphases on creativity, efficiency, ethics, authorship, and adoption.}
\label{tab:RQ}
{\small
\begin{tabularx}{\linewidth}{ll}
\toprule
Study & Research Questions / Objectives \\
\midrule
\makecell[tl]{\citep{french_creative_2023} French et al., 2023 \\\emph{Creative use of OpenAI} \\\emph{in Education}} & \makecell[tl]{\textbullet~``Describe and evaluate our experiences using AI tools
with BSc Games Programming \\undergraduates as part of their coursework''
\\\textbullet~``Provide [students] with opportunities for focused engagement with OpenAI that would\\ enhance their technical and problem-solving skills, as well as refine their abilities to analyse \\the current capability and potential of the technology, based on their own experiences''\\\textbullet~``Enhance communication skills in relation to AI technologies, moving away from \\social-media-framed hype and towards a more rigorous, well-informed perspective''\\\textbullet~
``Reduce students’ fears around AI replacing them in the future workplace''\vspace{0.2cm}}
\\
\makecell[tl]{\citep{grow_chatgpt_2023} Grow \& Khosmood, 2023 \\\emph{ChatGPT Game Jam}} & \makecell[tl]{\textbullet~Explore the potential of LLMs in game development by designing, convening, and \\evaluating a small ChatGPT Game Jam.\\\textbullet~Provide insights about the current capabilities and shortcomings of LLM-based \\game production.\vspace{0.2cm}}
\\
\makecell[tl]{\citep{lankes_ai-powered_2023} Lankes \& Stockl, 2023 \\ \emph{AI-Powered Game Design}} & \makecell[tl]{\textbullet~Investigate whether current AI chatbots can effectively support designers in specific \\design tasks, focusing on the professional perspective of game designer\\ \textbullet~Shed more light on the potential of AI chatbots like ChatGPT to support game design \\processes in general\vspace{0.2cm}} 
\\
\makecell[tl]{\citep{vimpari_adapt_or_die_2023} Vimpari et al., 2023 \\\emph{Adapt or Die:} \\\emph{TTIG in Game Dev}} & \makecell[tl]{\textbullet~``What are professionals' perceptions and attitudes towards TTIG systems and \\their future?''\\ \textbullet~``How are TTIG systems adopted and used in the creative practice now?''\\ \textbullet~``How will TTIG systems change and be used in future creative practice?''\vspace{0.2cm}}
\\
\makecell[tl]{\citep{begemann_gaidev_2024} Begemann \& Hutson, 2024 \\\emph{Empirical Insights} \\\emph{into AI-Assisted Game Dev}} & \makecell[tl]{\textbullet~Explore the potential of generative AI tools in revolutionising game development\\ pipelines by alleviating current limitations through a synergistic approach.\\\textbullet~Illuminate the pathways through which AI can streamline and enrich the game\\ development process, contributing to a broader discourse on the synergies between AI \\technologies and creative industries.\vspace{0.2cm}} 
\\
\makecell[tl]{\citep{boucher_resistance_2024} Boucher et al., 2024 \\\emph{Early Career Game Devs \& Gen AI}} & \makecell[tl]{\textbullet~``How, if at all, are (early game devs) using GAI in their development process?''\\ \textbullet~``What harms or benefits do they identify in using this technology?''\\ \textbullet~``How does their position as emerging professionals relate to their perceptions and \\usage of this technology?''\vspace{0.2cm}}
\\
\makecell[tl]{\citep{majgaard_pilot_2024} Majgaard, 2024 \\\emph{A Pilot Study:}\\ \emph{Engineering Students use GenAI}} & \makecell[tl]{\textbullet~``How can GenAI be used in the development of game-like applications for \\educational use?''\\ \textbullet~``How can we promote students' reflections on GenAI?''\vspace{0.2cm}}
\\
\makecell[tl]{\citep{panchanadikar_solodev_2024} Panchanadikar \& Freeman, 2024 \\\emph{I'm a Solo Developer}\\\emph{Indie Devs Online}} & \makecell[tl]{\textbullet~``What are the perceived novel opportunities and urgent risks of generative AI for \\indie game developers' efforts to innovate game development and production?''\\ \textbullet~``How can we design future generative AI technologies to enhance such \\opportunities and mitigate risks to better support these developers' efforts?''\vspace{0.2cm}}
\\
\makecell[tl]{\citep{shields_generatingtogether_2024} Shields et al., 2024 \\\emph{Generating Together:} \\\emph{Educational Visual Novel}} & \makecell[tl]{\textbullet~Understand how new generative AI technologies might integrate within video game \\development to support narrative and artistic productivity\\ \textbullet~Describe lessons learned from attempting to integrate LLMs and text-to-image models \\into the development of an educational visual novel\vspace{0.2cm}} 
\\
\makecell[tl]{\citep{alharthi_genaicreativity_2025} Alharthi, 2025 \\\emph{Generative AI in Game Design}} & \makecell[tl]{\textbullet~``How do game designers and developers perceive the value and potential of \\generative AI tools in their workflows?''\\ \textbullet~``How do generative AI tools influence creativity, productivity, and efficiency in \\game design and development?''\\ \textbullet~``What are the primary challenges and concerns game designers and developers face \\when using generative AI?''\vspace{0.2cm}}     

\end{tabularx}}
\end{table}

\section{Description of Included Studies}
%eMERGe item 8 - outcome of study selection%
\label{sec:results_included_studies}
%eMERGe item 10 - presenting characteristics of included studies%
%eMERGe item 11 - process for how studies are related%
We summarise the included studies across three tables: Tbl.~\ref{tab:RQ} outlines each study’s research questions/objectives; Tbl.~\ref{tab:dataQA} reports participants, data collection/analysis, duration, and CASP appraisal; and Tbl.~\ref{tab:aitype} situates them by type of game considered, \ac{GenAI} used, purpose, setting, scale of practice and geographic scope. To contrast adoption contexts, we split the literature according to their primary production purpose (\emph{Purpose}, Tbl.~\ref{tab:aitype}). Group A (\enquote{Learning}) comprises studies in learning and training environments, where \ac{GenAI} is used for skill development. Group B (\enquote{Production}) covers professional contexts, where \ac{GenAI} serves the production of public-facing assets or complete games. Moreover, we illustrate the study collection intervals together with key \ac{GenAI} milestones in Fig.~\ref{fig:timeline}.

\begin{table}[t]
\caption{Overview of participant numbers (N) and demographics, methods and duration of data collection in months (M), days (D), weeks (W) or not stated (---), methods of data analysis, and CASP appraisal weighting -- \enquote{Low} (L), \enquote{Medium} (M), \enquote{High} (H). }
\Description{Table describing demographics, sample sizes, methods, duration, analysis approaches, and CASP quality weighting for each study. Participants range from students and hobbyists to professionals; methods include interviews, case studies, surveys, and postmortems; analysis is mostly thematic or inductive.}
\label{tab:dataQA}
{\small
\begin{tabularx}{\linewidth}{lllllll}
& & & \multicolumn{2}{c}{Collection} & &\\
    \cmidrule(lr){4-5}
Study & Demographic & N & Methods & Duration & Analysis Methods & CASP\\
\midrule
\citep{french_creative_2023} & Students & 5 & Case studies & 9 M & Case study discussion & L \\[6pt]
\citep{grow_chatgpt_2023} & Hobbyists & 9, 2* & Online survey & 4 D & Case study discussion & L\\[6pt]
\citep{lankes_ai-powered_2023} & Industry Professionals & 3 & \makecell[tl]{Task-based assessment using ChatGPT; \\semi-structured interviews\\[4pt]} & --- & \makecell[tl]{Qualitative content \\analysis, inductive} & H\\
\citep{vimpari_adapt_or_die_2023} & Industry Professionals & 14 & \makecell[tl]{Semi-structured interviews; \\online pre-interview survey} & 2 M &  \makecell[tl]{Template analysis, \\inductive \\[4pt]} & H\\
\citep{begemann_gaidev_2024} & Academics & 2 & Diary/Logbook format & 10 D & Statistical analysis & M\\[6pt]
\citep{boucher_resistance_2024} & \makecell[tl]{Students, \\Aspiring Professionals} & 26 & \makecell[tl]{Semi-structured interviews; \\embedded researcher observation} & 11 W &  \makecell[tl]{Thematic analysis, \\inductive \& deductive\\[4pt]} & H\\
\citep{majgaard_pilot_2024} & Students & 36 & Student reports, classroom log notes & 20 W & \enquote{Grounded Theory} (?) & M\\
[6pt]\citep{panchanadikar_solodev_2024} & Industry Professionals & 3091* & Online post mining & 17 M &  \makecell[tl]{Thematic analysis, \\inductive} & H\\
[4pt]\citep{shields_generatingtogether_2024} & Industry Professionals           & 9                                       & Postmortem case study                                           & ---                  &  \makecell[tl]{Reflective analysis \\as case studies\\[4pt]}       & L            \\
\citep{alharthi_genaicreativity_2025} & Industry Professionals           & 42 / 9 *& \makecell[tl]{Online survey;\\ semi-structured interviews}                       & ---                  &  \makecell[tl]{Thematic analysis, \\inductive}              & M\\
\multicolumn{7}{l}{}\\
\multicolumn{6}{l}{* \citep{grow_chatgpt_2023}: 9 pre-survey and 2 post-survey ($\subseteq$ 9); \citep{panchanadikar_solodev_2024}: 3091 online posts; \citep{alharthi_genaicreativity_2025}: 42 online survey and 9 interview ($\subseteq$ 42).}
\end{tabularx}
}
\vspace{-0.5cm}
\end{table}

\subsection{Group A: Learning}
These four studies show how educational settings offer diverse perspectives on \ac{GenAI} for game production. \citet{boucher_resistance_2024} studied a US university summer programme simulating commercial game development with small, diverse intern teams. Its industry-readiness focus fostered debates on authorship and ethics alongside tool use, framed by the authors as resistance where scepticism toward \ac{GenAI} intertwined with questions of identity and ethics in emerging professional cultures. 
In a UK classroom, \citet{french_creative_2023} examined early integration of \ac{LLM} and \ac{TTIG} tools into individual game projects. Unlike the collaborative, industry-simulated setting of \citeauthor{boucher_resistance_2024}, this study highlighted how students negotiated technical limitations, with expectations defined more by tool deficiencies than transformative potential. \citet{majgaard_pilot_2024} provide a comparable classroom approach, studying Danish engineering students building serious games in small teams. Here, \ac{GenAI} was treated as an additional teammate whose output required constant verification. Shifting toward a production-oriented model, \citet{shields_generatingtogether_2024} documented a US serious games project where students and academics develop an educational visual novel. Like \citeauthor{boucher_resistance_2024}, the work simulated professional production but within academic constraints, with managing \ac{GenAI} assets proving as time-consuming as traditional art.

\subsection{Group B: Production}
These studies span a range of professionals settings, from cross-regional industry surveys to focused case studies, time-limited events, and large-scale analyses of community discourse. \citeauthor{alharthi_genaicreativity_2025} conducted a mixed-methods study with professionals from the MENA region, Europe, North America, and Asia. A survey %of 42 respondents 
identified ideation, asset creation, and programming as primary use-cases. Follow-up interviews revealed both enthusiasm and concern in terms of potential efficiency gains and erosion of creative authenticity when integrating \ac{GenAI} into established workflows. 
Where \citet{alharthi_genaicreativity_2025} mapped practices across multiple studios, \citet{begemann_gaidev_2024} turned inward to their own ten-day project. Acting as developers, their diary study focuses on the integration of \ac{TTIG} and 3D generation tools, highlighting a persistent gap in the production of high-quality 2D concept art and usable 3D assets. Studying a ChatGPT-based game jam, \citet{grow_chatgpt_2023} offer an even more time-constrained perspective. In contrast to previous pipelines, this context prioritised rapid prototyping and improvisation, allowing participants to explore unconventional uses of \ac{GenAI}. Post-event reflections revealed benefits of faster idea generation, but also a reduced sense of authorship. \citet{lankes_ai-powered_2023} explored similar uses of ChatGPT in longer-term indie studio practice. Interviews with three Austrian experts on design exercises reveal that, while \ac{GenAI} supported initial ideation, the outputs were not readily transferable to production without extensive refinement. 
While \citeauthor{lankes_ai-powered_2023} focused on a small number of experts, \citet{panchanadikar_solodev_2024} scaled outward to examine more than 3000 posts from international online developer communities. Discussions often portrayed \ac{GenAI} as a co-worker that could accelerate solo workflows, particularly for non-specialist tasks, while also producing errors that required close supervision. Concerns were also raised about asset quality, copyright issues, as well as over-reliance on the technology. The study further captured how community norms shaped this discourse, with enthusiasm for innovation tempered by peer-to-peer caution.
The first study of \ac{GenAI} in a specific industry, \citet{vimpari_adapt_or_die_2023} offered yet another perspective through interviewing professionals from Finnish game studios of different size on their use of \ac{TTIG}. Their study reveals both optimism about the technology's creative potential as well as unease over ethical challenges and commercial pressures to adopt new tools. The authors note the pragmatic framing of tensions, attempting to balance artistic ambitions with the realities of market competitiveness. 

\begin{table}[t]
\caption{Overview of study foci, identifying (1) type of game; (2) type of \ac{GenAI} used: Large Language Models (LLM), Text-to-Image Generation (TTIG), Image-to-3D (I23D) and Text-to-3D (T23D), Audio and Speech Generation (Aud. / Spe.); (3) primary production purpose; (4) setting; (5) scale of practice: solo, 2 to 11 people (S, small), 12 - 50 (M, mid-size); more than 50 (L, large); and (6) geographic
scope. Studies focused on a particular model are marked accordingly: C (LLM column) for ChatGPT and D (TTIG column) for DALL·E.
}
\Description{Table showing study focus across type of game (entertainment, serious, mixed), type of GenAI used (LLMs, text-to-image, 3D, audio/speech), purpose (learning vs production), setting, scale (solo, small, mid, large teams), and geographic scope. Coverage spans Europe, USA, MENA, and global contexts.}
\label{tab:aitype}
{\small
\begin{tabularx}{\linewidth}{lXccccXXXl}

    &      & \multicolumn{4}{c}{Type of GAI} & & & & \\
    \cmidrule(lr){3-6}
Study                                   & Game Type & \adjustbox{angle=90}{LLM} & \adjustbox{angle=90}{TTIG} & \adjustbox{angle=90}{T23D / I23D} & \adjustbox{angle=90}{Aud. / Spe.}
& Purpose & Settting & Scale & Geographic Scope                \\
\midrule
\citep{french_creative_2023} & Entertainment & C & D & & & Learning & Classroom & Solo & UK\\
\citep{grow_chatgpt_2023} & Entertainment & C &   & & & Production & Event Based & Solo & USA\\
\citep{lankes_ai-powered_2023} & Mixed & C & & & & Production & Industry & S & Austria\\
\citep{vimpari_adapt_or_die_2023} & Entertainment & & $\times$ & & & Production & Industry & S, M, L & Finland\\
\citep{begemann_gaidev_2024} & Entertainment & & $\times$ & $\times$ & & Production & Research & S & USA\\
\citep{boucher_resistance_2024} & Entertainment & $\times$ & $\times$ & & & Learning & Industry & S & USA \\
\citep{majgaard_pilot_2024} & Serious & $\times$ & $\times$ & & $\times$ & Learning & Classroom & S & Denmark\\
\citep{panchanadikar_solodev_2024} & Mixed & $\times$ & $\times$ & & $\times$ & Production & Industry & Solo, S & Global\\
\citep{shields_generatingtogether_2024} & Serious & $\times$ & $\times$ & & & Learning & Research & S & USA\\
\citep{alharthi_genaicreativity_2025} & Mixed & $\times$ & $\times$ & & $\times$ & Production & Industry & S, M& MENA, Europe, NA, Asia    
\end{tabularx}
}
\end{table}

\section{Translation}
\label{sec:translation}
\begin{figure}[t]
    \centering
    \includegraphics[width=0.8\textwidth]{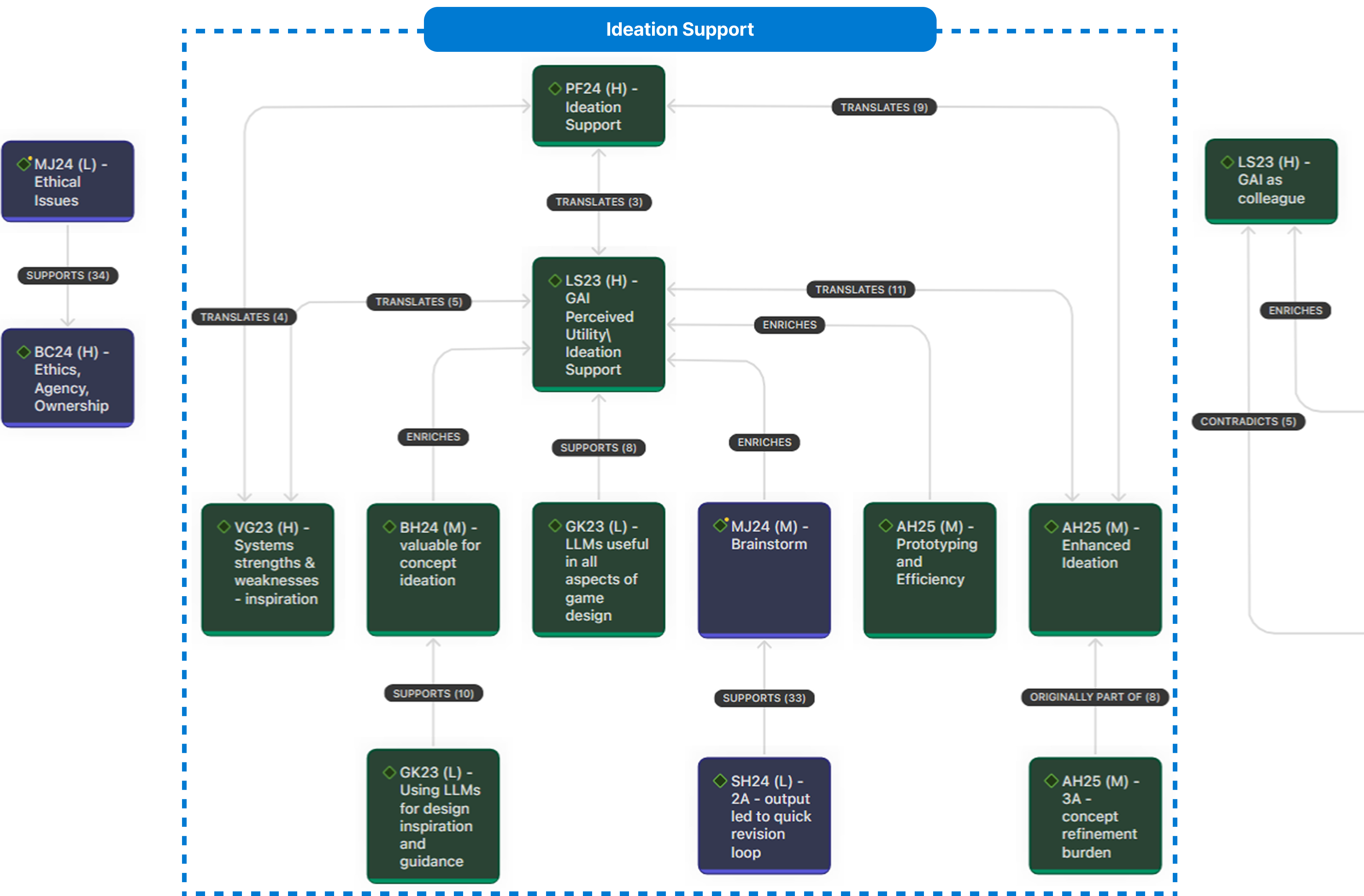}
    \caption{Exemplary translation cluster from the full synthesis map (Supplementary Materials) with 
    the \enquote{ideation support} interpretation at the top. The hues signal that most interpretations stem from the \Pgroup{Production} group with only two contributing  from the \Lgroup{Learning}.}
    \Description{A network diagram showing the translation cluster for “ideation support.” Central nodes from production and learning studies are connected by Translates, Supports, and Enriches relations. Most nodes are dark green (production), with two purple nodes (learning). The cluster is enclosed with a dashed blue line denoting the overall theme of the clustered interpretations. The structure illustrates how diverse studies converge on ideation as a shared claim while adding different contexts and conditions.}
    \label{fig:clusterzoom}
\end{figure}

%eMERGe item 13 - process of translating studies%
\paragraph{Procedure} Similarly to \citet[pp.~55-57]{campbell_evaluating_2011}, we (a) applied reciprocal translation in two stages (first within, and then across the Learning and Production groups) and (b) adopted a chronological procedure with an \enquote{index} paper: using the oldest primary study as the index against which other were compared \citep[p. 47, 65]{campbell_evaluating_2011}, we proceeded in publication order, adding one paper at a time and iteratively updating the synthesis. In the rest of the paper, we use \enquote{scope} to denote the level or focus of a 2\textsuperscript{nd}-order interpretation (e.g. whether a finding refers to asset inclusion, pipeline efficiency, or adoption attitudes). Interpretations were related using a custom vocabulary of connectors in Atlas.ti:
\begin{description}
    \item[Translates] Conceptually equivalent at the same scope.
    \item[Supports] Adds further first-order interpretations of the same type.
    \item[Enriches] Introduces a new dimension to an existing concept.
    \item[Contradicts] Opposes another interpretation at the same scope.
    \item[Originally part of] Retains an author-defined interpretive hierarchy from the primary source.
\end{description}
Fig.~\ref{fig:clusterzoom} illustrates their use. The first four relationships connect nodes from different primary studies, and the last enables us to distinguish original hierarchies from those developed through the synthesis process. Relations were assigned collaboratively, with details provided in Appx.~\ref{app:reflexivity}. Once each group’s network was internally connected, they were brought together for cross-group translation using the first four types of annotators. Where the scope was ambiguous, we returned to the primary texts to verify the interpretation, giving greater scrutiny to lower-rated studies (Tbl.~\ref{tab:casp-summary}). The final integrated network retained all relation types, with group-specific and cross-group structures remaining visible for subsequent synthesis (Supplementary Materials).

%eMERGe item 12 - outcome of relating studies%
\paragraph{Outcome} Relating the studies yielded a dense \Pgroup{Production} network (79 unique interpretations, 105 links), a more compact \Lgroup{Learning} network (37 unique interpretations, 35 links), and a cross-group layer (24 links). From here onwards (incl. figures), colour-code each of these primary groups as exemplified above. Across all three spaces, most relations either \emph{Enrich} or \emph{Support} an existing interpretation rather than asserting strict equivalence (\emph{Translates}). In \Pgroup{Production} over half of the links (54\%) \emph{enrich} the linked interpretation with new conditions or mechanisms, while 18\% \emph{support} the same claim with further 1\textsuperscript{st}-order interpretations. Only 12\% \emph{translate} directly, and \emph{contradictions} are very rare (1.9\%). In \Lgroup{Learning}, the pattern shifts slightly towards corroboration, with \emph{Supports} (49\%) roughly matching \emph{Enriches} (43\%) and a similar scarcity of \emph{Translates} and \emph{Contradicts}. This scarcity is not surprising, given the overall modest number of primary studies and the HCI research tendency not to replicate but complement existing work. %It indicating that studies more often build upon rather than accurately duplicating or opposing one another.

Similarly, cross-environment links are dominated by \emph{Enriches} (83\%), with very few  \emph{Translates} and \emph{Contradicts}. This pattern reflects that when \Lgroup{Learning} and \Pgroup{Production} connect, they add contextual detail but, as of now, do not identify fully equivalent phenomena across domains. Where contradictions occur, they often mark boundaries in framing, e.g.~\Lgroup{ \enquote{LS23 - GAI as colleague}} vs.~\Pgroup{\enquote{BC23 - AI as Tool}} reflecting differences in role conceptualisation between environments, which is expanded on in our Discussion (\ref{sec:discussion}). 

%eMERGe item 14 - outcome of translation%
In the resulting relational map, Each node retained the verbatim 2\textsuperscript{nd}-order interpretation from its source, linked to the underlying 1\textsuperscript{st}-order data in Atlas.ti for direct traceability. Fig.~\ref{fig:clusterzoom} shows a close-up of the \enquote{ideation support} translation cluster, illustrating how the relational patterns identified above play out in practice. Links to \Pgroup{\enquote{AH25 - Prototyping and Efficiency}} and \Pgroup{\enquote{VG23 - Systems strengths \& weaknesses - inspiration}} enrich the core idea by adding mechanisms and context-specific detailing, whilst links such as  \Lgroup{\enquote{MJ24 - Brainstorm}} support the claim by providing concrete examples from the classroom. Contributions to the capsule come from 8 primary sources (LS23, GK23, etc.).

\section{Synthesis Procedure}
\label{sec:synthesis-procedure}
%eMERGe item 15 - synthesis process%
The cross-group relational map from reciprocal translation (Sec.~\ref{sec:translation}) forms the basis for our synthesis. Clusters were used as the primary organisational unit, and \emph{Translates} and \emph{Contradicts} relations formed our comparative baseline, establishing where studies aligned or diverged before considering how interpretations were extended or substantiated. For each connection, we re-read the primary study extracts and recorded memos on scope, framing, and (dis-)agreement. We then examined \emph{Enriches} connections to understand how core interpretations were extended, re-contextualised, or conditionally bounded, and \emph{Supports} to locate where arguments gained density through 1\textsuperscript{st}-order evidence. 

We then turned these observations into provisional narratives for each cluster, noting where interpretations bridged clusters or sat at their margins. Where an interpretation connected to multiple clusters (different colours in synthesis map), allocation was based on contextual fit, with decisions reviewed collaboratively (details in Appx.~\ref{app:reflexivity}). This process yielded nine \emph{synthesis capsules}, representing our nascent 3\textsuperscript{nd}-order interpretations, each corresponding to one or a merger of several clusters. Each capsule brings together the interpretations, conditions, corroborations, and scope limits of one phenomenon. Short identifiers (e.g., \Lgroup{C1.2}, \Pgroup{C8.3}) for each underlying interpretation will be used consistently with the colour-coding in the remaining text to indicate where individual 2\textsuperscript{nd}-order interpretations sit within the synthesis, and to maintain linkage with the 1\textsuperscript{st}-order evidence. The resulting \enquote{synthesis map}, an update of the original \enquote{translation map}, is available in the Supplementary Materials. Capsule labels and narratives express our 3\textsuperscript{nd}-order interpretations, which were then refined into the following line-of-argument synthesis, linking the nine clusters into a coherent account of how \ac{GenAI}'s impact has been studied across learning and production settings. 

\section{Outcome of Synthesis}
\label{sec:outcome-synthesis}

%eMERGe item 16 - outcome of synthesis process%
We complement a \emph{line-of-argument synthesis} (Sec.~\ref{sec:methods}) with brief paragraphs of \emph{implications and recommendation}. The structure is provided by our 3\textsuperscript{rd}-order interpretations, corresponding to the nine identified capsules (\textbf{C1}-\textbf{C9}) that each capture a distinct but connected aspect of \ac{GenAI}'s impact. These span production-level practices, socio-technical positioning, adoption dynamics, governance, and authorship consequences, presented in an order that surfaces both their internal logic and their cross-linkages. Parenthetical code references (e.g., \Lgroup{C1.2}, \Pgroup{C8.3}) point to the respective capsule (e.g.~\textbf{C1}, \textbf{C8}) and individual interpretations extracted previously; colours follow the evidence-group scheme introduced in the translation stage (Sec.~\ref{sec:translation}). Via the full codebook (Supplementary Materials), each interpretation can be traced back to its primary study and CASP score. Interpretive decisions during the synthesis were cross-checked collaboratively (details in Appx.~\ref{app:reflexivity}), with alternative framings compared against  and grounded in the primary studies.

\subsection{C1: Human-in-the-loop refinement as the production norm}
\paragraph{Synthesis} Generative outputs operate as provisional artefacts which require expert intervention before they can be accepted for production. The studies reviewed are consistent in this observation, and that correction is integral to the work rather than a marginal contingency (\Pgroup{C1.1}, \Lgroup{C1.2, \Lgroup{C1.3}}). In professional practice, both indie developers and those from bigger studios describe iterative prompting as a central mechanism for recovering creative intent and reconciling the model's tendencies with the stylistic and functional requirements of the project (\Pgroup{C1.4}, \Pgroup{C1.5}). Studies in educational settings convey a similar stance, albeit with different objectives: \acp{LLM} are positioned as programming tutors and scaffolds rather than autonomous generators, and in-engine runtime code generation during development is constrained to relatively simple tasks (\Pgroup{C1.6}, \Lgroup{C1.14}, \Lgroup{C1.15}). Students' accounts of routinely editing outputs before they can be applied further highlight the normality of such intervention (\Lgroup{C1.16}). Early stage ideation and its benefits are further discussed in \textbf{C2}. In code-based tasks, practitioners engage in iterative exchanges with the system in which error messages are used diagnostically to structure subsequent prompts, allowing gradual convergence on a working solution (\Pgroup{C1.7}). In image generation, stylistic stability is sensitive to asset class, with backgrounds proving more variable than character designs, and acceptable results are more often the outcome of selective curation across multiple outputs than of single, high-fidelity generations (\Lgroup{C1.8}, \Lgroup{C1.9}). Even when outputs are close to acceptable, preparatory work such as training and asset conditioning remains time-intensive and requires expertise (\Lgroup{C1.10}). The need for refinement is even stronger in 3D production, where unedited outputs are rarely suitable for direct inclusion due to technical and stylistic tensions with project constraints (\Pgroup{C1.11}). How these constraints appear at engine boundaries is detailed in \textbf{C4}, and the organisational controls that stabilise inclusion are set out in \textbf{C8}. At this juncture, human editorial decision-making determines which variants anchor subsequent work, which deviations from intent are acceptable, and which corrective strategies preserve alignment with the project’s overall requirements (\Pgroup{C1.12}, \Pgroup{C1.13}). The data thus repositions prompting from the simple act of query formulation to a form of progressive specification work in which the developer controls the range of acceptable solutions while maintaining coherence across the project. No primary evidence suggests that \ac{GenAI} can produce inclusion-ready artefacts without human revision. Differences between studies concern the efficiency of refinement loops (\textbf{C3}) and the extent of adoption (\textbf{C6}), but these operate at distinct analytical scopes and do not contradict the claim that human intervention remains necessary at the point of inclusion. 

\paragraph{Implications and Recommendations} To capture actual working practices, \ac{GenAI} evaluation should extend from static measures of output quality to capturing characteristics of the refinement process, including time-to-acceptable-edit, the latency with which errors become visible, and the degree of control available in resolving them (\Pgroup{C1.7}–\Pgroup{C1.11}). Reporting should be disaggregated by asset class and pipeline stage to make evident where acceptance rates are higher or lower, and to inform targeted technical and organisational interventions (\Lgroup{C1.8}–\Pgroup{C1.11}). Tool design should support accountable refinement by preserving edit histories, enabling side-by-side comparison of variants, and recording the rationale for inclusion decisions (\Pgroup{C1.12}, \Pgroup{C1.13}). In education, assessments might place greater value on refinement competence (i.e. the capacity to diagnose, steer, and integrate outputs) rather than overemphasising the quality of first-attempt generation (\Pgroup{C1.6}, \Lgroup{C1.14}–\Lgroup{C1.16}). While the evidence base is strongest for code and 2D imagery, findings for 3D outputs, although convergent, are sparse and corresponding claims should be interpreted as indicative  (\Pgroup{C1.11}).

\subsection{C2: Early-stage ideation scaffold, not autonomous authorship}
\paragraph{Synthesis} Across all settings, the primary value of \ac{GenAI} in game development lies in their contribution to ideation rather than in any capacity for autonomous authorship, described as routine ideation support (\Pgroup{C2.1}, \Pgroup{C2.2}) or enhanced ideation (\Pgroup{C2.3}). Studio reflections further reinforce this premise, emphasising the utility of generators for broadening the range of ideas prior to human selection (\Pgroup{C2.4}). In game jams, \acp{LLM} are used explicitly for design inspiration and problem-solving guidance (\Pgroup{C2.5}). Ethnographic analyses offer greater specificity about where generative inspiration is reliable and how it is enacted. Users articulate system strengths and weaknesses in relation to inspiration (\Pgroup{C2.9}), and engage in exploratory, playful routines within an established creative process (\Pgroup{C2.7}, \Pgroup{C2.8}). Educational settings provide convergent evidence, showing that \ac{GenAI} can produce narrative variety that expands the range of options available for curation (\Lgroup{C2.6}). These accounts suggest that ideation is most effective under prototyping regimes where speed, disposability, and low-fidelity exploration are prioritised (\Pgroup{C2.10}). The productivity of such early-stage work is enhanced by explicit role design, with systems positioned as creative assistants within collaborative workflows that maintain clarity about human and machine responsibilities (\Pgroup{C2.11}, \Pgroup{C2.12}). In classroom settings, this is complemented by reports of brainstorming and knowledge gains that accelerate team progress (\Lgroup{C2.13}, \Lgroup{C2.14}), as well as explicit role allocations that preserve human judgement in key creative decisions (\Lgroup{C2.15}). A single counterpoint in the corpus tempers any tendency to overgeneralise from ideation capacity to autonomous authorship. Arguably little evidence from jam practice shows that producing a complete game from a single prompt remains difficult (\Pgroup{C2.16}). This reinforces that \ac{GenAI}'s ideation function is bounded: human framing, evaluation, and integration are constitutive of the process.

\paragraph{Implications and Recommendations} The implication for both research and practice is that evaluation criteria should be tailored to the ideation role. Systems should be assessed on the quality and diversity of options surfaced, their ability to support constraint-aware suggestion, and the degree to which they enable clarity in partner and workflow roles. Analyses should avoid conflating early-stage variety with later production efficiency; where efficiency does occur, it is likely to be context-specific and is examined in \textbf{C3}. These ideation outputs are provisional and require subsequent refinement for inclusion; the conditions and costs of that refinement are addressed in \textbf{C1}.

\subsection{C3: Efficiency claims contested}
\paragraph{Synthesis} Across the corpus, efficiency is frequently asserted, yet the evidence shows it to be contingent upon development phase, asset class, and practitioner expertise. Expert accounts describe time saving in everyday practice (\Pgroup{C3.1}), and review-level analyses identifies resource efficiency and faster completion for bounded activities (\Pgroup{C3.2}, \Pgroup{C3.3}). Classroom reports note faster progress but only in specific exercises(\Lgroup{C3.5}), while solo developers report advantages of focused, cost-effective development (\Pgroup{C3.11}). Many of these reports arise in early stages, as noted in \textbf{C2}. Studio reflections place the value of acceleration in rapid prototyping and concept ideation, but only under conditions of sufficient AI expertise to make use of the system productively (\Pgroup{C3.9}, \Pgroup{C3.10}). Review evidence clarifies that such advantages apply to specific tasks rather than to any wholesale acceleration of the development pipeline (\Pgroup{C3.2}, \Pgroup{C3.3}). In this reading, efficiency emerges as a conditional and situated property that depends on the alignment between the scope of the task, the model’s affordances, and the practitioner’s capability to configure, prompt, and integrate outputs effectively. Realised gains also depend on integration fit and organisation controls, which are detailed in \textbf{C4} and \textbf{C8}. The corpus also documents clear limits to efficiency claims: art teams report that image generation did not accelerate the production of background assets at the point of inclusion (\Lgroup{C3.6}), and model preparation remains time-consuming even where generation is rapid (\Lgroup{C3.7}). In controlled classroom exercises, manual editing proved faster than generative assistance when the specification was tight (\Lgroup{C3.8}). At the scale of individual labour, configuration, prompting, and verification overheads can erode net benefits, reframing some time-saving claims as speculative investments that do not consistently pay off (\Pgroup{C3.12}).
These findings support a qualified conclusion in which efficiency is not a general property of \ac{GenAI} in game development but a phase and context dependent outcome. Gains are most likely in early-stage work where disposability and rapid iteration are valued, and least likely in downstream asset inclusion and preparation where refinement and integration work, discussed further in \textbf{C1}, reabsorb much of the time apparently saved upstream. 

\paragraph{Implications and Recommendations} Evaluation should be conducted at the scale of discrete activities rather than entire pipelines, with reporting that pairs elapsed time with rework avoided and verification burden. Expertise prerequisites should be made explicit, and efficiencies in prototyping should not be generalised to production contexts that are contradicted by evidence from art and preparation stages. Evidence is strongest for  prototyping and educational micro-tasks, while findings from downstream contexts are more consistently neutral or negative. Measured effects are further shaped by pipeline constraints and governance arrangements, treated in \textbf{C4} and \textbf{C8}.

\subsection{C4: Pipeline, integration, and artefact constraints}
\paragraph{Synthesis} Across settings, present tools are reported to misalign with game-production pipelines, which require packaged, versioned artefacts, traceable provenance, and engine-conformant formats. Weak evidence from a game jam setting suggests that model outputs are optimised for text interaction rather than asset integration, producing a modality mismatch that necessitates adapters, post-processing, and human evaluation before inclusion (\Pgroup{C4.1}). Classroom settings support such affordance gaps more strongly, e.g. highlighting that debugging via prompts is difficult when errors need targeted inspection within an engine context (\Lgroup{C4.2}). Similarly, the absence of standardised transparency support creates barriers for straightforward importing of outputs into art pipelines (\Lgroup{C4.3}). Unity coding with ChatGPT still presupposes prior programming competence, which sustains reliance on existing expertise rather than enabling turnkey integration (\Lgroup{C4.10}). Unity runtime code generation has been found useful only for a narrow set of 3D primitives (\Lgroup{C4.11}). Sensitivities to labelling and training data, and to inference parameters, surface as practical hazards. For example, higher values for the \ac{LLM} temperature parameter produce more diversity but also nonsensical output that requires detection and curation (\Lgroup{C4.12}–\Lgroup{C4.13}). Jam participants add that generated code comments and rationales are often not optimal for explanation or hand-off, which would limit integration especially in bigger projects (\Pgroup{C4.14}). Students also describe hallucination as a routine risk to be managed (\Lgroup{C4.15}). In art pipelines, stability is contingent on asset class, with background images exhibiting greater instability than character assets and therefore demanding more corrective work prior to import (\Lgroup{C4.4}). Studio reflections consolidate these observations into process-level requirements. Teams report that current technology requires refinement to meet industry standards, that tool selection materially affects output quality, and that the generation of 3D models remains challenging for production purposes (\Pgroup{C4.5}, \Pgroup{C4.8}–\Pgroup{C4.9}). The studios respond by formalising integration as a designed and testable process: dedicated evaluation methods are required at integration gates to institutionalise model choice, acceptance thresholds, and regression checks over time (\Pgroup{C4.6}). Usability and integration dependencies are identified as system conditions that mediate realised value (\Pgroup{C4.7}). The surrounding infrastructure provides the organisational means to manage misfits and variation, as explored in \textbf{C8}.
No data supports frictionless, end-to-end integration. Apparent successes in practice remain local and depend precisely on the adapters, expertise, and evaluation infrastructures documented above, which means they do not refute the underlying constraint. Thus, without structured post-processing, explicit acceptance criteria, and accountable checks, generated materials fail to satisfy pipeline requirements across both code and art contexts (C4.1–\Lgroup{C4.10}, C4.13). Claims about time saved should be read with the task-level analysis in \textbf{C3} in mind.

\paragraph{Implications and Recommendations} We recommend to treat integration as a first-class design problem. Specifications should state asset-class acceptance criteria and error budgets; adapters for compositing and import, including transparency handling and structured outputs, should be engineered as part of the workflow rather than improvised ad hoc. Debugging affordances need to extend beyond dialogue, with facilities for inspection, tracing, and deterministic reproduction within the engine. Integration gates should be supported by evaluation harnesses and provenance capture at hand-off, so that model updates and parameter changes are auditable. Progress ought to be reported as reductions in post-processing effort, hallucination incidence, and re-integration time, rather than as claims of seamless substitution. In short, value accrues when integration work is made visible and improvable within the pipeline, not when generation alone is optimised without regard to the artefact constraints that ultimately govern inclusion (see \textbf{C8} for further discussion on governance).

\subsection{C5: Socio-technical positioning: assistant, colleague, partner}
\paragraph{Synthesis} Practitioners often describe \ac{GenAI} in game development as a collaborator-like aide that augments rather than replaces human creativity. Expert developers report engaging with these systems much as they would with other  colleagues, coordinating critique, exploration, and turn-taking (\Pgroup{C5.1}). For independent developers, the same stance appears in how they characterise the system as a \enquote{co-worker} and through collaboration routines that structure feedback cycles and sustain momentum in small teams (\Pgroup{C5.2}, \Pgroup{C5.3}). Across the entire industry, we can generalises this pattern as creative augmentation, making explicit that the function of the technology is to enhance the creative process rather than to automate it (\Pgroup{C5.5}). In learning contexts, students position the system as a teammate, integrating it into small-group coordination and using it to support collective progress (\Lgroup{C5.6}). Whether this stance becomes routine depends on adoption patterns by role and setting, examined in \textbf{C6}. The durability of this collaborator stance depends on deliberate socio-technical configuration. Treating \ac{GenAI} as social entities allows to prescribe persona design and conversational conventions that align model behaviour with human workflow needs (\Pgroup{C5.4}). These mechanisms underpin the assistantship role, enabling it to generalise across contexts while maintaining accountability to human judgement. They also help explain variation in metaphor use: where the interaction is scaffolded with social design features, colleague and partner framings become more acceptable and robust. Professional identity and craft accountability, however, set clear boundaries on personification. Industry ethnography shows that some practitioners reject colleague metaphors for \ac{GenAI}, positioning the system as a tool when authorship, skill recognition, and responsibility are at stake (\Lgroup{C5.7}). This refutation does not challenge the augmentation premise itself; rather, it specifies the contexts in which certain social metaphors are used. \ac{GenAI} as assistants forms a baseline across both industry and learning settings, with colleague and partner roles remaining contingent on the absence of other conflicts. Further concerns about legitimacy and labour protections that shape these limits are discussed in \textbf{C7}.

\paragraph{Implications and Recommendations} For both research and practice, the implication is to design for assistantship as the durable baseline. Systems should provide controls for persona and tone, make turn-taking and critique affordances visible, and link generative contributions to artefact-level justifications. Evaluation of collaboration features should consider how well they support human-led decision making and preserve craft accountability. Where colleague metaphors are desired, their social design needs to be explicit, as in the persona-based approaches of \textbf{C5.4}, and paired with attribution and provenance measures discussed in \textbf{C9}, as well as with protections for ownership and credit addressed in \textbf{C7}. In this way, interactional convenience is supported without blurring authorship or diminishing professional responsibility.

\subsection{C6: Access, democratisation, and adoption}

\paragraph{Synthesis} \ac{GenAI} is reported to lower entry barriers to game development, widening participation while revealing uneven patterns of adoption. % and role-specific sentiment. 
Independent participants describe a democratising effect, with tools motivating and enabling a broader range of people to engage in game making (\Pgroup{C6.1}, \Pgroup{C6.2}). Expert developers express optimism about the potential for adoption (\Pgroup{C6.7}), and classroom settings reveal detailed accounts of practical on-ramps, enumerating uses such as code generation, dialogue agents, Unity runtime code and procedural content generation, and real-time creation (\Lgroup{C6.18}–\Lgroup{C6.23}). Industry settings align these gains with perceived efficiency in prototyping regimes (\Pgroup{C6.9}, \Pgroup{C6.12}). The persistence of participation gains depends on fit and skill. Studio reflections locate sustained use where AI expertise is present and benefits accrue in concept generation (\Pgroup{C6.15}–\Pgroup{C6.16}), while asset-class boundaries (e.g. 3D models) temper early optimism (\Pgroup{C6.14}). Ethnographic accounts document adoption as a situated process embedded within specific communities, routed through assessments of system strengths and weaknesses, ease of use, and evolving perceptions of system capabilities and development trajectories (\Pgroup{C6.5}, \Pgroup{C6.11}, \Pgroup{C6.38}–\Pgroup{C6.39}, \Pgroup{C6.41}–\Pgroup{C6.44}). Perception shifts are shaped both by direct interaction with the tools and by the socio-economic and organisational contexts in which they are deployed. Content qualities also influence immediate utility, with overly verbose output cited as degrading day-one usefulness (\Lgroup{C6.46}).
Constraints on diffusion are visible in both review and classroom evidence, which describe adoption as limited or experimental rather than widespread (\Pgroup{C6.4}, \Lgroup{C6.17}). Industry ethnography records predominant scepticism in craft-intensive roles where human skill, authorship, and accountability are central to identity (\Lgroup{C6.8}). Concerns about originality and displacement also shape acceptance thresholds, conditioning the willingness to adopt rather than denying access outright (\Pgroup{C6.6}, \Pgroup{C6.10}). Studio cases further show that friction in integrating outputs into pipelines can slow or stall adoption, even where initial enthusiasm is high (\Pgroup{C6.14}–\Pgroup{C6.16}). The combined picture is one of conditional expansion: entry barriers fall, but sustained adoption depends on alignment with role requirements, pipeline compatibility, practitioner expertise, and evolving perceptions.

\paragraph{Implications and Recommendations} Democratisation should be treated as a conditional process rather than a guaranteed outcome. Analyses should distinguish between access and sustained adoption, measuring uptake at the granularity of role, asset class, and pipeline stage. On-ramps such as those identified in the uses above require targeted investment in expertise development and integration fit if they are to translate into durable practice. Perception trajectories should be monitored and addressed, with attention to content qualities such as verbosity that may erode utility. Strategies for adoption need to incorporate and address originality and labour concerns so that widened access converts into stable and legitimate participation, rather than remaining a transient increase in experimentation (see \textbf{C9} on authorship and \textbf{C7} on legitimacy and labour protections).

\subsection{C7: Risks, ethics, and labour precarity}
\paragraph{Synthesis} Accounts of opportunity are accompanied by recognition of ethical uncertainty and the exposure of creative labour. The primary studies identify ethical and legal ambiguity, concerns about originality, and the possibility of labour displacement as salient systemic issues (\Pgroup{C7.1}–\Pgroup{C7.3}). Practitioner perspectives translate these into risk categories that guide human oversight and the establishment of authorship boundaries. In solo and independent practice, these include career growth-, personal investment-, creativity-, and intellectual ownership risk. Each of these frames potential losses in professional advancement, the erosion of authorship, the disputability of ownership and credit, and the narrowing of net benefits through sunk costs in skill acquisition and verification (\Pgroup{C7.4}–\Pgroup{C7.7}). Studies in studio settings reveal parallel concerns in the form of job security anxieties, legal uncertainties, and disputes over data ownership (\Pgroup{C7.8}–\Pgroup{C7.10}).
Ethnographic evidence further specifies how such risks are enacted and sustained in practice. Anticipatory displacement appears as a lived experience, shaping community-level narratives and leading some practitioners to strategically avoid integrating tools into workflows where job loss is perceived as a credible threat (\Pgroup{C7.11}). In solo contexts, personal investment risk manifests in the transfer of costs from task execution to configuration, verification, and skill development (\Pgroup{C7.5}), which in turn determines the realisation of nominal efficiency gains (\textbf{C3}). Across settings, uncertainty about originality and attribution function as a gate on asset inclusion, with adoption dependent on whether provenance and credit can be formalised (\Pgroup{C7.2}, \Pgroup{C7.7}; see \textbf{C9} for authorship policy).
The corpus contains no scope-matched evidence that contradicts the existence of ethical or labour risk. Where counterpoints appear, they operate at different scopes: augmentation framings in \textbf{C5} focus on the mechanics of collaboration without disputing the need for legitimacy measures, while adoption optimism in \textbf{C6} captures attitudes that coexist with ownership, legal, and job-security constraints. It  therefore seems reasonable to say that without safeguards for authorship and role protection, the benefits of \ac{GenAI} in game development remain contested and fragile.

\paragraph{Implications and Recommendations} Risk management should be treated as a primary design and governance concern. Provenance capture and attribution policies need to be embedded in workflows to address originality concerns (\Pgroup{C7.2}), alongside the codification of licensing and data-ownership practices (\Pgroup{C7.9}–\Pgroup{C7.10}). Role and asset-specific acceptance criteria should be defined to preserve the visibility of human authorship and protect the stability of junior labour (\Pgroup{C7.4}, \Pgroup{C7.8}). Anticipated personal investment costs should be budgeted explicitly, both in training provision and in evaluation time (\Pgroup{C7.5}). Such practices create the conditions under which technical benefits can be realised without normalising precarity or undermining the creative legitimacy of the work. These safeguards underpin the collaboration stances in \textbf{C5}, influence adoption patterns described in \textbf{C6}, and align with medium-aware authorship measures in \textbf{C9}.

\subsection{C8: Governance of practice and workflow friction}

\paragraph{Synthesis} Across settings, realised value from \ac{GenAI} is mediated by explicit workflow design, evaluation criteria, and organisational infrastructure rather than generic model capability. In game jams and rapid prototyping, organisers position \acp{LLM} inside the process, with guidance that formalises when to prompt, how to iterate, and where to hand off so that outputs enter subsequent work coherently (\Pgroup{C8.1}). Industry  evidence suggests the same conclusion at organisational scale, diagnosing immature or absent infrastructure that constrains institutional adoption even where local enthusiasm is high (\Pgroup{C8.2}). Studio reflections consolidate governance as a first-class requirement by specifying the responsibilities that make integration possible in practice, including methods for evaluating AI at integration gates, model selection policies, acceptance thresholds per asset class, regression checks across pipeline stages, and quality assurance hand-offs that preserve accountability (\Pgroup{C8.3}). Where these structures are weak or absent, professional ethnography records friction in workflow that suppresses day-to-day uptake despite perceived promise, indicating that organisational design is a condition for routine use rather than a post hoc enhancement (\Lgroup{C8.4}).
These accounts link directly to the pipeline and artefact constraints identified earlier in C4. Governance provides the organisational means by which misfits are managed and variation is controlled. Process placement prevents unreal expectations of end-to-end automation and clarifies the social and technical roles involved in prompting, critique, selection, and inclusion. Evaluation harnesses and acceptance criteria translate abstract capability into fit-for-purpose judgements at the level of asset class and pipeline stage. Regression checks and provenance capture make model and parameter changes auditable over time, which is necessary for stable collaboration across teams. In this framing, positive local results are not counter-examples to the need for governance; rather, they are instances where strong process design has already been installed.

\paragraph{Implications and Recommendations} Governance should be treated as a design object in its own right. Teams should specify role and asset-class acceptance criteria, implement evaluation harnesses and provenance capture at hand-off, and invest in integration infrastructure and team learning so that know-how is not confined to a few experts. Success should be measured not only in artefact quality and throughput but also in observed reductions in workflow friction and in the stability of inclusion decisions over time (\Pgroup{C8.1} - \Lgroup{C8.4}). In short, governance is the condition under which \ac{GenAI} can be incorporated accountably and at scale, rather than an optional layer added after capability has been demonstrated.

\subsection{C9: Authorship, materiality, and aesthetic consequence}

\paragraph{Synthesis} \ac{GenAI} is framed not only as a tool or collaborator but as a medium whose adoption redistributes authorship and shapes style. Industry-facing analysis frames the system as material, locating agency in craft decisions about how the medium is manipulated (\Lgroup{C9.1}). At the field level, review evidence identifies aesthetic flattening, in which reliance on generative outputs risks stylistic convergence that diminishes expressive distinctiveness (\Pgroup{C9.2}). Professional discourse situates these concerns within questions of ethics, agency, and ownership, making clear that authorship and credit become contested when AI-generated material is incorporated into final artefacts (\Lgroup{C9.3}). Ethnography underlines this contestation, with disagreement over whether prompting, selection, and post-editing constitute creative labour (\Pgroup{C9.4}). Divergence also appears across role boundaries: artists and programmers display different epistemic commitments to authorship, which shape their attribution practices and policy preferences (\Lgroup{C9.5}).
Practitioners also articulate futures in which legitimate, medium-aware use is possible without eroding distinctiveness. \enquote{Imagining beyond resistance} describes design-oriented strategies that reconcile adoption with stylistic diversity, including provenance-aware workflows, dataset curation, and editorial constraint setting (\Lgroup{C9.6}). Interactional practices such as anthropomorphising the system can support collaboration by making its behaviour more predictable in dialogue, although such personification does not resolve the underlying questions of authorship or credit (\Pgroup{C9.7}). The combined evidence indicates that treating \ac{GenAI} as a medium directly affects attribution norms and aesthetic outcomes.
The dataset contains no similarly scoped evidence that denies the link between medium and authorship or the reality of style effects. Points of tension occur in other domains: collaboration framings in \textbf{C5} address the interpersonal stance adopted towards the system, while risk and labour discussions in \textbf{C7} address legitimacy and protection. In both cases, these do not contradict the central claim that medium choice and use conditions directly impact creative credit and stylistic consequence.

\paragraph{Implications and Recommendations} Authorship policies must be medium-aware. Definitions of creative contribution in AI-mediated pipelines should be explicit, specifying thresholds for authorship claims across roles and artefact types, as well as mandating provenance capture and attribution procedures (see \textbf{C7}). Evaluation of \ac{GenAI} should extend beyond throughput to include measures of expressive distinctiveness, with style-diversity and convergence diagnostics used to detect and mitigate aesthetic flattening (\Pgroup{C9.2}). Where AI is incorporated, medium-aware practices such as those described in \Lgroup{C9.6} should guide dataset selection, constraint setting, and documentation of human editorial labour. While interactional personification can support workflow fluency (\Pgroup{C9.7}), it must not substitute for clear, enforceable authorship and credit rules that protect the integrity of creative work. Where these policies meet pipeline realities, practical arrangements are discussed in \textbf{C4} and \textbf{C8}.

\section{Discussion}
\label{sec:discussion}
%eMERGe item 17 - summary of findings%

We briefly summarise our findings w.r.t.~our four research questions (Sec.~\ref{sec:introduction}), referring back to the original sections where additional summarisation would create too much redundancy.

\paragraph{RQ1} Our overview of included studies (Sec.~\ref{sec:results_included_studies}) and the corresponding Table  \ref{tab:RQ}--\ref{tab:dataQA} inform \textbf{\enquote{how existing qualitative studies differ}}. We identified two broad study contexts: educational and professional/industry-related. Within the first, researchers have studied the exploration of \ac{GenAI} as part of game development classroom exercises and professional training. The latter in contrast comprised studies of game professionals in studios from indie to AAA, but also developers not clearly affiliated with industry as present in game jams. The education-focused studies involve students or early-career interns (US/UK/Denmark) and mixed student teams (US), while the professional studies report data from broad, cross-regional industry samples and, more rarely, with a national (Finland, Austria) focus. The majority of works studies entertainment games wit more than half of all studies focused on industry professionals. The identified studies were methodologically diverse with interviews and online surveys only accounting for half the studies. Crucially though, more than half of studies show medium to strong methodological weaknesses as articulated through \ac{CASP} weightings. 

\paragraph{RQ2} It is the function of our synthesis (Sec.~\ref{sec:outcome-synthesis}) and the corresponding map (Supplementary Materials) to identify how findings \textbf{\enquote{characterise the ways in which \ac{GenAI} is adopted, used, and perceived within game development workflows}} (RQ2) in the primary literature. We do so by integrating the studies' 2\textsuperscript{nd}-order interpretations into 3\textsuperscript{rd}-order, interpretive constructs that examine how reported concepts qualify, extend, or sit in tension with one another and what shared patterns they jointly imply \cite{campbell_evaluating_2011,malpass_medication_2009}. We identify nine overarching themes (3\textsuperscript{rd}-order interpretations) in total, spanning refinement practices, ideation and efficiency, pipeline and artefact constraints, socio-technical positioning, access and adoption, labour and legitimacy, governance, and authorship (\textbf{C1}–\textbf{C9}). Across \textbf{C1}–\textbf{C4}, the synthesis characterises the use of \ac{GenAI} as human-in-the-loop and front-loaded within the workflow: prompting operates as a form of progressive specification work, generative outputs function as provisional artefacts that require expert curation and editing before inclusion, and efficiency appears as a conditional, phase- and asset-dependent property that is most credible in upstream ideation and prototyping and least so at the point of integration, where refinement and reconciliation with engine and pipeline constraints are concentrated. It is tempting to first conceive the human-\ac{GenAI} co-creative act as \emph{task-divided creativity}, where \enquote{partners take specific roles within the co-creative process [e.g. prompting, generating], producing new concepts satisfying the requirements of one party} \citep{ kantosalo2016modes}. However, the evidence suggests that it is better understood as \emph{alternating co-creativity}, in which the \enquote{co-creative partners take turns in creating a new concept satisfying the requirements of both parties}. Here, we understand \ac{GenAI}’s \enquote{requirements} in the sense of generative bias or output constraints. Across \textbf{C5}, \textbf{C6} and \textbf{C8}, adoption emerges as a form of conditional democratisation rather than straightforward diffusion. Entry barriers fall and new on-ramps into game making are created, yet sustained uptake is anchored where \ac{GenAI} can be cast as an assistant that fits existing craft identities, role responsibilities, asset classes and locally available integration competence, instead of displacing these outright. Read through Schön's distinction between reflection-on-action and reflection-in-action, the material reveals both modes and clarifies how such assistantship is maintained in practice. Participants' retrospective accounts of game jams, classroom projects and studio work describe how using \ac{GenAI} led them to re-evaluate design and development decisions, including whether its involvement had been beneficial and how it ought to be positioned in future projects (reflection-on-action). Their descriptions of iterative prompting, diagnosis and revision in situ, by contrast, exemplify reflection-in-action as they adjust prompts, constraints and inclusion thresholds in response to the system's behaviour and to pipeline-specific breakdowns~\citep{schon1983reflective}. Taken together, these observations indicate, first, that reflection-in-action has become a routine part of the micro-practices through which practitioners manage generative bias and keep provisional outputs compatible with engine and artefact requirements (\textbf{C1}–\textbf{C4}); and second, that reflection-on-action underpins the longer-term calibration of when, where and for whom \ac{GenAI} is considered appropriate, feeding into role-specific judgements about assistantship, adoption and governance (\textbf{C5}, \textbf{C6}, \textbf{C8}). Across \textbf{C7}–\textbf{C9}, perceptions are structured by questions of legitimacy and medium: ethical and labour risks, concerns about originality and ownership, and anxieties about stylistic convergence act as gates on acceptable practice and are negotiated through provenance and attribution arrangements, workflow-level governance and medium-aware authorship policies. At the level of RQ2, the synthesis thus portrays \ac{GenAI} in contemporary game development as a provisional, assistant-like technology whose principal value lies in human-steered ideation and exploratory work, and whose longer-term integration is conditioned by pipeline fit, organisational design and contested settlements around labour, authorship and aesthetic distinctiveness.

\paragraph{RQ3} Our reciprocal translation (Sec.~\ref{sec:translation}) and the final synthesis map (Supplementary Materials) show \textbf{\enquote{in what ways findings across studies converge, complement, or contradict one another}}. Across the corpus, findings converge on four points: (1) \ac{GenAI}'s present value is upstream (ideation) rather than end-to-end authorship,(2) human intervention remaining necessary for output integration, (3) efficiencies are conditional rather than general, and (4) realised benefit is dependant on pipeline fit and existing governance. They complement one another by situating these claims in different environments: classrooms and game jams illustrate how \ac{GenAI} is taken up for experimentation and team creativity, while professional accounts highlight asset-specific constraints, organisational dependencies, and risks to authorship and labour. Direct contradictions are rare; when they do appear, they mark boundaries or limitations: reports of prototyping speed-ups contrasting with evidence of slowdowns during asset inclusion, or classroom framings of \ac{GenAI} as a \enquote{teammate} versus professional accounts that position it as a tool when authorship or responsibility is at stake. These divergences refine rather that overturn the shared conclusions by underlining the conditions under which claims about value and use can be sustained.

\paragraph{Tensions}
Several productive tensions emerged from our translation across settings and analytic scopes. These did not constitute refutational contradictions in the sense of \citeauthor{noblit_meta-ethnography_1988}, as they rarely addressed the same claim at the same analytic level. We therefore do not claim to have conducted a refutational analysis and present these tensions here in a concentrated fashion and as complement to our reciprocal translation. Overall, the tensions reveal how assumptions about \ac{GenAI} shift as practices move between exploratory, educational, and professional environments. 
First, tensions around \emph{efficiency} signal that speed is not a stable property of \ac{GenAI} use but depends on where in the development arc the system is applied. Reports of rapid prototyping and early-stage acceleration (\textbf{C2}; \Pgroup{C3.1}; \Lgroup{C3.5}; \Pgroup{C3.11}) sit alongside evidence that downstream inclusion slows once refinement and debugging are required (\textbf{C1}; \Pgroup{C3.6}–\Pgroup{C3.8}; \textbf{C4}). While these findings do not contradict each other directly, they underline that efficiency is phase-dependent. What appears as acceleration in exploratory phases becomes mitigated by downstream integration costs, thus creating a boundary between speed of ideation and production throughput.
Second, we find tensions in the \emph{social positioning} of \ac{GenAI}. Classroom and game jam settings describe the technology as a teammate or partner supporting their collective momentum (\Lgroup{C5.6}; \Lgroup{C2.13}–\Lgroup{C2.15}), whereas several professional settings revert to a tool-based framing when authorship or craft identity are at stake (\Pgroup{C5.7}; \Pgroup{C7}). This reframing is not simply linguistic. It reveals that the social ontology of \ac{GenAI} is shaped by local norms of responsibility. When learning and experimentation are the primary goals, relational framings flourish. When reputation or liability are implicated, boundaries around authorship become more guarded. Third, studies diverge in expectations of what \ac{GenAI} can \emph{legitimately produce}. Educational contexts emphasise breadth and exploratory flexibility (\Lgroup{C2.4}-\Lgroup{C2.6}), while professional accounts constrast this with the practical realities of asset preparation and integration (\textbf{C4}; \Pgroup{C1.8}–\Pgroup{C1.11}). The tension is therefore not a disagreement about capability, but a contextual difference in what counts as acceptable. In exploratory contexts, rough concepts are sufficient; in production, viability is governed by specific constraints that sharply limit which outputs can progress.
Finally, contrasts in \emph{adoption narratives} show that \ac{GenAI} is interpreted through the lens of local labour conditions. Independent and student developers often characterise \ac{GenAI} as a means of lowering barriers to experimentation (\Lgroup{C6.1}–\Lgroup{C6.2}; \Lgroup{C6.18}–\Lgroup{C6.23}). Industry accounts, by contrast, foreground concerns around authorship and the precarious position of junior practitioners (\Pgroup{C7.4}–\Pgroup{C7.10}). This friction illustrates that enthusiasm and caution are not opposites but reflections of different forms of exposure to risk. For some, \ac{GenAI} expands opportunities; for others, it complicates the conditions under which creative work is made accountable.
Taken together, these disagreements, whilst not sufficient to perform refutational translation, offer insight into where claims about \ac{GenAI} are stable and where they are contingent. They show that speed, collaboration, capability, or adoption are not intrinsinc qualities of the systems but relational outcomes that depend on the structure of the environment in which they are embedded. They also highlight opportunities for future HCI research to focus on the situated arrangements that shape what \ac{GenAI} becomes in practice. 

\paragraph{RQ4} Our comparison of studies, translation and synthesis all inform \textbf{\enquote{which conceptual, empirical, or methodological gaps remain}} as starting point for informing future research priorities. The following gaps and recommendations complement those identified in Sec.~\ref{sec:outcome-synthesis} and \ref{sec:contextualisation}.

\emph{Conceptually}, studies often conflate short-term access with durable adoption (\textbf{C6}), and pipeline-level efficiency with task-level speed-ups (\textbf{C3-C4}). Refinement is widely observed (\textbf{C1-C3}) but rarely theorised in its own right, making it difficult to distinguish basic prompting from the broader competence required to diagnose or steer outputs. Similarly, framings of \ac{GenAI} as assistant, colleague, or tool (\textbf{C5-C7}) remain context-dependent but under-specified, making it unclear under which conditions each stance is most productive. 

\emph{Empirically}, many of our primary studies draw on only small groups of game makers. Even taken together, the combined sample remains very limited and poorly representative of the global game industry \citep{kerr2017global}. Most existing studies capture the early(ish)-adoption phase of this transition (Fig.~\ref{fig:timeline}), and we still lack a clear view of the large-scale structural changes that \ac{GenAI} pipelines may introduce to game development. Related, empirical coverage is uneven: 2D imagery and code are relatively well evidenced, while 3D, animation, audio, and mixed-pipeline artefacts remain under-studied (\textbf{C1, C3, C4}). Engine-level integration (\textbf{C4}) and organisational governance (\textbf{C8}) are widely discussed but rarely examined in situ with direct observation or instrumentation. No studies so far are of longitudinal nature, which could e.g. inform insights on adoption w.r.t.~retention, abandonment, or skill accumulation (\textbf{C6}). While insights into industry studies are of high relevance, these could be complemented with studies on hobbyists and game-jams which are presently rare and \enquote{underpowered}. Surprisingly, studies with a focus on US industry as important game development ecosystem so far only focus on small studios or solo developers. Ideally, this should be complemented with insights from bigger studio and compared to findings from other local economies to contrast local industry practices.

\emph{Methodologically}, reporting practices vary substantially, with some studies clearly separating pipeline stages and asset types, while others aggregate them, limiting comparability (\textbf{C1, C3, C4}). In addition to this, details of model choice and configurations are not consistently provided, making it difficult to trace how specific claims were grounded (\textbf{C7, C9}). Effects on ideation are usually described narratively rather than evaluated with explicit criteria. Related to the identified gaps on empirical coverage, researchers must become more transparent about who was interviewed and the exact timing of the data collection and analysis. In the case of \ac{GenAI} adoption, even a few months can lead to very different findings. Game industry professionals are accustomed to highly flexible ways of working, often planning only two months ahead \citep{kultima2015developers} and being extremely sensitive to timing and to trends circulating in the field. More broadly, sample sizes, durations, and units of analysis vary markedly across studies which impacts comparability (Fig.~\ref{fig:timeline}). This particularly held for case studies and postmortems; we consider these valuable, but only if complemented with future studies of similar design and potentially more participants. Diary studies have so far only been employed to capture the observation of solo academic designers; we highlight these as excellent vehicle to gain the longitudinal insights promoted earlier and believe that deployment with professionals and hobbyists could yield particularly valuable insights. More generally, short data collection intervals for distinct research questions in Fig.~\ref{fig:timeline} illustrate the need for more longitudinal research. Our detailed \ac{CASP} rating (Supplementary Materials) informs further methodological improvements. Here, we only note the arguably most prominent shortcoming: in several studies, qualitative coding was done at an insufficient standard, e.g.~omitting descriptions of 2\textsuperscript{nd}-order interpretations or grounding in the raw data. This makes consolidation efforts such as ours particularly hard and should be urgently avoided.

\section{Contextualisation in Game Production Trends}
\label{sec:contextualisation}
In game development, a five-year observation window is still relatively short. We are working with a phenomenon in flux, where new tools and pipelines are tested and adopted while their impacts unfold. At the same time, AI is by no means new to the industry. From early procedural content generation to machine learning for character control, games have long experimented with automation \citep{yannakakis2018artificial}. \ac{GenAI} represents a significant step forward, yet its integration into production pipelines will take time. Procedural methods (Sec.~\ref{sec:background}) illustrate this: despite decades of research and  demonstrations, their use remains uneven and often inefficient within mainstream development environments. 

Studying \ac{GenAI} in isolation risks misinterpretation, since many of the phenomena currently discussed align with long-standing dynamics in game development. Asset production, for example, has always required varying rounds of iteration \citep{kultima2015developers, kultima2018game}, whether assets are produced in-house or outsourced. Generative tools may alter the tempo or form of these loops, but they may not eliminate the fundamental need for revision and reworking. Similarly, the apparent democratising potential of \ac{GenAI} continues decades of progress in the accessibility of development environments. Unity’s shift towards free licensing models already broadened participation and changed industry practices \citep{kerr2017global, young2021unity}, while low-code tools such as GameMaker, Bitsy, and Ren'Py have enabled game-making for those unfamiliar to programming. Yet all platforms also encode biases in the kinds of games they most easily produce \citep[e.g.][]{consalvo2021reading}. However, practitioners also often subvert these conventions by misusing or stretching tools \citep{hugill2013creative}. \ac{GenAI} similarly risks reproducing existing norms rather than enabling radical departures, though practitioners may also find ways to push against its constraints.

Another important dimension is that the central problems of game development are increasingly less about technology and more about human resources. Post-mortem studies have shown that difficulties in production often stem from soft skills, team coordination, and organisational dynamics rather than mere technical obstacles \citep{politowski2021game}. In this respect, \ac{GenAI} may intersect with challenges of communication and collaboration as much as with hard-skills related efficiency or automation.
The industry’s structural precarity must also be acknowledged: long before and in parallel to \ac{GenAI}, volatility was created by shifting business models and changing technologies \citep{kultima2018game,bulut2020precarious}, keeping up precarity and uncertainty. The rise of free-to-play, for instance, forced many premium studios to adopt new pipelines and revenue strategies \citep{kerr2017global}. The recent post-pandemic wave of lay-offs continues this pattern of instability. Certain creative roles, such as concept artists, writers, or voice actors, among other roles have always been fragile. \ac{GenAI} therefore enters a labour landscape already marked by insecurity, and its effects need to be considered against this broader background.

Seen in this light, \ac{GenAI} appears less a radical rupture than a continuation of existing trajectories. Its use so far remains largely conservative, producing incremental variations in ideas and styles. While it can multiply the quantity of assets, characters, levels, or environments,  \enquote{more} content does not necessarily translate into more engaging or meaningful experiences. Audience tastes are themselves conservative, reinforced by industry consolidation and mass-market imperatives, although counter-trends such as the rise of K-pop and other non-Hollywood media suggest that shifts in cultural consumption are always possible. The studies synthesised here highlight that a significant impact is underway, but sustainable insights on \ac{GenAI}'s long-term impact require more time, further study and stronger contextualisation in the broader cultural, economic, and technological dynamics of the industry. 

%eMERGe item 18 - strengths, limitations, reflexivity%
\section{Reflexivity \& Study Limitations}
\label{sec:limitations-reflexivity}

As an interdisciplinary team working across game HCI, computational creativity, and design practice, our backgrounds inevitably informed how we interpreted the primary studies. We approached the topic from a broadly agnostic stance toward \ac{GenAI}, recognising its potential value when used to complement rather than replace human creative practice, but also acknowledging potential threats, e.g to environment, individuals and society. To complement individuality in interpretation, all stages of our meta-ethnography were conducted collaboratively, with iterative comparison of alternative readings and repeated reference back to the primary sources. This process ensured that our 3\textsuperscript{rd}-order interpretations remained grounded in the evidence and not in our preconceptions. We present a full account of reflexive procedures, including stage-by-stage details of our collaborative workflows and positionality statements in Appx.~\ref{app:reflexivity}.

While we emphasise the transformative impact of \ac{GenAI}, it is important to recognise that several other forces are reshaping daily work and production in game industry. Hybrid work models, the current geopolitical climate, and declining consumer purchasing power all influence how studios can structure and sustain the workforce. In addition, the long-running rise of third-party tools and game engines continues to democratise production and accelerate change, and this trend predates \ac{GenAI}. While we contextualise our findings in empirically grounded game production trends (Sec.~\ref{sec:contextualisation}), our synthesis does not integrate the interaction of the identified \ac{GenAI} phenomena with these wider developments at the same level as within \ac{GenAI}. Research and policy must take a holistic stance.

Moreover, it would have been desirable to ground our findings game development theory specifically. Crucially though, such theory is largely absent: we lack a comprehensive understanding of game development practices and the regional differences in how developers think and work \citep{kerr2017global, cadin2006hrm, khaled2023generative}. When studying game development, it is essential to situate the data not only in time but also in place. Practices vary from project to project, but they also differ across local cultures, companies, and levels of experience \citep{kultima2018game}. The present synthesis cannot account for these differences in sufficient detail to directly support theory building, but our overview of studies (Sec.~\ref{sec:results_included_studies}; \ref{sec:translation}) and identified research gaps (Sec.~\ref{sec:discussion}) point at specific contexts to investigate further with the appropriate methods and epistemologies to further the development of such theory. At present, we ground our findings in theory that is applicable beyond games.

The quality of our synthesis rests heavily on the quality of its underlying primary studies. The evidence base remains modest and uneven across contexts, with primary studies varying in conceptual depth and methodological quality. This asymmetry shapes the density of our translation networks and limits the precision with which comparisons can be made across the corpus. Poor 2\textsuperscript{nd}-order interpretations in some studies required us to augment such interpretations to foster comparability across studies. This introduces an additional interpretative layer. 

Fig.~\ref{fig:timeline} illustrates that the qualitative studies underpinning our synthesis are overall modest in numbers and scattered over more than three years with typically short data collection intervals (\citet{panchanadikar_solodev_2024} exceptionally collected social media data over a longer timeframe) and distinct research questions. This situation did not allow for our synthesis to track changes over time. Instead, it combines several points in time, each representing potentially transient phenomena. Therefore, the synthesis must not be understood as a snapshot of the present. Moreover, interactions with other developments in games (see above) matter and the present moment is unusually opaque -- many practitioners describe the current landscape as \enquote{dark}, shaped by overlapping and rapidly changing pressures: data regulation, declining consumer spending, AAA instability, consolidation, and intensifying global competition, among others. This uncertainty makes it difficult to forecast how \ac{GenAI} pipelines will actually take hold. 

On a similar note, while our translation explicitly compares different contexts, the synthesis may obfuscate that that game development is by no means a unified or standardised practice. Game products vary dramatically in complexity, scope, and form, and so do game development contexts and communities. Communities of practice (e.g. AAA, mobile and hyper-casual, arthouse, academic, hobbyist, etc.) overlap, but they also remain distinct in their norms, expectations, and structural conditions. Additionally, each sector operates under different constraints and (e.g. commercial) incentives. A consequence of abstracting from evidence drawn from various development contexts, our synthesis highlights mutual support and tensions in identified phenomena, e.g.~between production and learning environments, but we remind our audience of the diversity in practices from which our evidence was sourced. 

While we hope for our findings to provide guidance for e.g.~practice or policy, it is for the reasons above that we strongly recommend to not use the synthesis insights in isolation; instead, we suggest to use them as a starting point which, complemented with the identified research gaps and quality appraisal, can inform the collection of further in-the-moment data to inform definite decisions or forecast future developments. Sensitive to the decision task, this research may have to balance an holistic approach with sensitivity to e.g.~diversity in practice or local characteristics.

\section{Conclusion}
%eMERGe item 19 - recommendations & conclusions%
We conducted the first qualitative research synthesis (QRS) on the impact of \ac{GenAI} on videogame production. We employed PRISMA-S to systematically search the literature, identifying a corpus of 10 primary literature items to then expose to a meta-ethnography as popular QRS method, guided by the eMERGe framework and supported by CASP quality appraisal. Through reciprocal translation, line of argument synthesis and visual mapping, we identified and summarise nine overarching themes in existing qualitative work. Together with our detailed comparison of qualitative studies along various axes such as research goals, settings, demographic and methodology, we highlight gaps in research and provide recommendations for future work. Since the impact of \ac{GenAI} does not happen in a vacuum, we contextualise our findings within broader trends in game production. We hope that these insights will benefit industry stakeholders, researchers and policymakers with grounded and concentrated insights to guide future practice, research and support governance.

\begin{acks}
We highlight individual contributions using the Contributor Roles Taxonomy \citep[CRediT,][]{credit}: Funding acquisition (CG), Conceptualisation (CG), Methodology (AT, CG), Investigation (AT, AD, JC, CG), Formal analysis (AT, AD, CG),  Visualisation (AT, JC), Validation (AT, AD, JC, CG), Supervision (CG, AD),  Writing - original draft (ALL), Writing - revised draft (ALL). 
\end{acks}

\appendix

\section{Search Strategy}
\label{app:search}
Search strings were constructed using Boolean operators, with terms within each block combined using \texttt{OR} and blocks joined using \texttt{AND}. Wildcards (e.g., \texttt{*}) were used to capture morphological variation, and quotation marks were used to ensure semantic specificity in multi-word terms.\\

\noindent\textbf{Finalised multi-block query (general form):}
\begin{spverbatim}
("game dev*" OR "game creat*" OR "game prod*" OR "game design" OR "game studio*" OR "game pipeline*" OR "game jam*" OR "game hackathon*")
AND
("gen* AI" OR genAI OR "AI gen*" OR "generative machine-learning" OR "generative model*" OR "foundation model*" OR "large language model*" OR LLM* OR ChatGPT OR "text-to-image generat*" OR TTIG OR LTGM OR "transformer-based generat*" OR "transformer model*" OR "diffusion-based generat*" OR "diffusion model*" OR "Stable Diffusion" OR Midjourney OR DALL*E OR DALL-E* OR DALL·E* OR DALLE* OR "multimodal generative model*")
AND
(qualitative OR interview* OR survey* OR "focus group*" OR questionnaire* OR "case stud*" OR "thematic analysis" OR "practice-based" OR "design research" OR "research through design" OR ethnograph* OR autoethnograph* OR "grounded theory" OR "diary study")
\end{spverbatim}

\noindent\textbf{Scopus query:}
\begin{spverbatim}
( TITLE-ABS-KEY ( "game dev*" OR "game creat*" OR "game prod*" OR "game design"
OR "game studio*" OR "game pipeline*" OR "game jam*" OR "game hackathon*" ) )
AND
( TITLE-ABS-KEY ( "gen* AI" OR "genAI" OR "AI gen*" OR "generative machine-learning"
OR "generative model*" OR "foundation model*" OR "large language model*" OR llm*
OR "ChatGPT" OR "text-to-image generat*" OR "TTIG" OR "LTGM"
OR "transformer-based generat*" OR "transformer model*"
OR "diffusion-based generat*" OR "diffusion model*" OR "Stable Diffusion"
OR "Midjourney" OR "DALLE*" OR "DALL-E*" OR "DALL E*" OR "DALL·E*"
OR "multimodal generative model*" ) )
AND
( TITLE-ABS-KEY ( qualitative OR interview* OR survey* OR "focus group*"
OR questionnaire* OR "case stud*" OR "thematic analysis" OR "practice-based"
OR "design research" OR "research through design" OR ethnograph*
OR autoethnograph* OR "grounded theory" OR "diary study" ) )
\end{spverbatim}

\begin{table}[t]
\label{tab:searchdomain}
\small
\centering
\caption{Search terms grouped by query domain for structured literature search (transposed).}
\Description{Table showing the structured literature search strategy. Search terms are grouped into four domains: game-related terms, generative AI terms, creativity and collaboration concepts, and qualitative/mixed-methods research approaches.}
\begin{tabularx}{\linewidth}{X X X X}
\toprule
\textbf{Game domain} & \textbf{GenAI domain} & \textbf{Concept domain} & \textbf{Methodology domain} \\
\midrule
game dev*            & gen* AI                        & creativ*                & qualitative \\
game creat*          & genAI                          & computational creativity & interview* \\
game prod*           & AI gen*                        & creative AI              & case stud* \\
game design          & generative machine-learning    & AI creativity            & thematic analysis \\
game studio*         & generative model*              & AI for creativ*          & practice-based \\
game pipeline*       & foundation model*              & collaborat*              & design research \\
game jam*            & large language model*          & co-creat*                & research through design \\
game hackathon*      & LLM*                           & mixed-initiative*        & ethnograph* \\
                     & ChatGPT                        & ideat*                   & autoethnograph* \\
                     & text-to-image generat*         & feedback loop*           & grounded theory \\
                     & TTIG                           & enhanced creativ*        & diary study \\
                     & LTGM                           &                          & \\
                     & transformer-based generat*     &                          & \\
                     & transformer model*             &                          & \\
                     & diffusion-based generat*       &                          & \\
                     & diffusion model*               &                          & \\
                     & Stable Diffusion               &                          & \\
                     & Midjourney                     &                          & \\
                     & DALL·E (all variants)          &                          & \\
                     & multimodal generative model*   &                          & \\
\bottomrule
\end{tabularx}
\end{table}

\subsection*{Comparative Query Evaluation}

To assess the relevance and impact of the fourth, more abstract block (conceptual framings and interaction modalities), we compared the output of a three-block query (blocks 1–3) against the full four-block query. The comparison revealed that while the fourth block filtered out non-relevant technical papers, it occasionally excluded conceptually relevant but less explicitly framed studies. Consequently, we included results from both configurations, manually reviewing overlaps to ensure a balanced and inclusive study set that captured both technically and conceptually grounded work.

\section{Reflexive Positioning and Analytic Collaboration}
\label{app:reflexivity}
This appendix expands on how reflexivity was embedded throughout the synthesis, including (1) an overview of collaboration across all stages of the meta-ethnography and (2) brief positionality statements from each author. It complements the brief reflexivity notes in Sec.~\ref{sec:limitations-reflexivity} and seeks to foster transparency in interpretive qualitative synthesis.

\subsection{Collaborative Analytic Process Across Meta-Ethnography Stages}
Our collaborative workflow followed the seven stages of meta-ethnography (Sec.~\ref{sec:methods-meta-ethnography}), with reflexive discussion integrated throughout. We met weekly or biweekly basis with additional asynchronous exchanges via Slack and shared documents. Author D contributed additional contextualisation in game industry practices (Sec.~\ref{sec:introduction} and \ref{sec:contextualisation}) post-synthesis.

\paragraph{\textbf{(1) Selecting meta-ethnography and defining focus (Sec.~\ref{sec:introduction};\ref{sec:methods})}}

The initial scope, research questions, and methodological rationale were drafted by authors B, C, and E and refined in group discussions. Reflexivity centred on clarifying our assumptions about GenAI's role in creative practice and ensuring that these would not pre-structure the synthesis.

\paragraph{\textbf{(2) Determining what is relevant (Sec.~\ref{sec:selection-eligibility}; \ref{sec:search_strategy})}}

Eligibility criteria were jointly developed at the project outset. Authors A-C and E participated in paper screening, with disagreements resolved consensually in group discussions. Agreement on inclusion was sought at both abstract/title screening and full-text review stages. Iterative reflections concerned our scoping of `game development' and how boundaries around creative roles may influence interpretation.

\paragraph{\textbf{(3) Reading included studies (Sec.~\ref{sec:study_selection_selection_process})}}

Author A read all included studies and authors B, C and E subsets. Notes and preliminary observations were shared in group meetings. Reflexive discussion focused on how our disciplines shaped what we each attended to in the text (e.g., workflow details, creative reasoning, organisational conditions).

\paragraph{\textbf{(+) Qualitative Research Appraisal (Sec.~\ref{sec:execution-qa})}}
The appraisal of each paper was done collaboratively between authors A and B on a shared spreadsheet with the modified 11-question CASP checklist \citep{long_optimising_2020}. The appraisers rated each criterion and provided a detailed justification resting on evidence from the source material (Supplementary Materials). Reflexivity concerned the tipping-point criteria and fair treatment of studies with different goals and methodology.

\paragraph{\textbf{(4) Determining how studies are related (Sec.~\ref{sec:execution-extraction})}}

The relationships between concepts were mapped collaboratively. Authors A and E extracted the interpretations from the primary sources and drafted initial affinity structures, which were then reviewed and reworked jointly with authors B and C. Regular meetings were used to challenge interpretive assumptions and explore alternative conceptual groupings.

\paragraph{\textbf{(5) Translating studies into one another (Sec.~\ref{sec:translation})}}

Reciprocal translations were developed iteratively. Semantic connectors were initially assigned individually by authors  A-C and E and then validated collaboratively until consensus was reached. Cases of uncertainty or disagreement were discussed in group review sessions and 1\textsuperscript{st}- and 2\textsuperscript{nd}-order interpretations were consulted when discrepancies arose. This stage involved explicit reflexive questioning about what was being emphasised or omitted in our comparisons, and how CASP ratings influence a study's coverage. 

\paragraph{\textbf{(6) Synthesising translations (Sec.~\ref{sec:synthesis-procedure})}}

Authors A-C and E created 3\textsuperscript{rd}-order interpretations over several synthesis meetings. Competing interpretations were discussed and alternative framings (e.g., authorship, accountability, reflective practice) tested against the primary studies. Decisions were grounded in textual evidence, not majority opinion.

\paragraph{\textbf{(7) Expressing the synthesis (Sec.~\ref{sec:outcome-synthesis})}}

The line-of-argument was drafted collaboratively and iteratively revised by authors A-C and E. Reflexive attention was given to the risks of over-generalisation, the influence of our shared agnostic stance toward GenAI, and the need to clearly surface tensions and boundary conditions within the literature.

\subsection{Individual Author Positionality}
The following brief statements outline the epistemic and disciplinary perspectives each author brought to the synthesis. These statements are intentionally anonymised, focusing on intellectual positioning rather than biographical detail.

\begin{description}
    \item[Author A] is a designer and researcher, with a background in product and service design, currently working within HCI, with a focus on how creative practitioners think, decide, and collaborate with technology. They approach \ac{GenAI} as both a design material and a workplace actor, interested in how its integration changes expertise, authorship, and everyday design work.
    \item[Author B] works in game HCI, with a focus on player and developer experiences, and the dynamics of tool adoption in game production. They approach GenAI pragmatically, attending to how it reorganises labour, coordination, and decision-making within teams.
    \item[Author C] specialises in computational creativity and interactive media research. Their work examines how creators engage with computational tools, and how such tools foster exploration, iteration, and creative identity. They adopt a broadly constructivist stance, viewing creativity as emerging from interactions among people, artefacts, and technological systems.
    \item[Author D] %studies game development practices using qualitative, ethnographic, and industry-engaged methods. Their work examines production cultures, design processes, and how teams negotiate technology adoption in real-world contexts. They bring sensitivity to organisational dynamics, team coordination, and the lived realities of creative labour, foregrounding how GenAI interacts with professional identity, craft, and studio workflows.
    has done extensive research on game-development practices, particularly the lived experiences of developers working in an evolving landscape. They have engaged in conversations with hundreds of developers from a wide range of nationalities and roles, from leaders in AAA studios to independent creators and hobbyists with diverse design values. Focusing on Northern Europe, the author’s interactions also include developers from across East and South Asia, the Americas, Australia, and South Africa. They bring sensitivity to organisational dynamics and the lived realities of creative work.
    \item[Author E] researchers computational creativity, human-AI collaboration, and game AI, with a longstanding interest in the boundaries between human and machine creative agency. They seek to support a societally and culturally sustainable future of AI via empirical work, and by letting the latter guide AI research. Their perspective foregrounds reflective practice and the ethical and epistemic implications of integrating AI into creative processes. Here, they employed an interpretative stance and a critical mindset toward \ac{GenAI} research and adoption practices. 
\end{description}

While individual perspectives differ slightly in criticality, we share a broadly agnostic stance toward \ac{GenAI}: recognising its potential value when used to complement, rather than replace, human creative work, and cautious of framings that overstate automation or diminish human judgement and creative expression. These shared commitments informed our interpretive focus on creative negotiation, reflective engagement, authorship, and accountability.

\bibliographystyle{ACM-Reference-Format}
\bibliography{bibliography}

\end{document}